\documentclass[fleqn,12pt]{article}

\usepackage{graphics,epsfig,cite}

\def\beq   {\begin{equation}}
\def\eeq   {\end{equation}}
\def\beqd  {\begin{displaymath}}
\def\eeqd  {\end{displaymath}}
\def\beqaa {\begin{eqnarray}}
\def\eeqaa {\end{eqnarray}}

\def\PR              {P_R^{}}
\def\PL              {P_L^{}}

\def\tW              {\t_W}

\def\rzw             {\sqrt{2}}

\def\noi {\noindent}

\def\ti  {\tilde}

\def\st  {\ti t}
\def\sb  {\ti b}

\def\sn  {\ti \nu}
\def\sl  {\ti \ell}

\def\nt  {\tilde\chi^0}

\def\chm {\tilde\chi^-}

\def\a   {\alpha}
\def\b   {\beta}
\def\t   {\theta}

\def\sz{\ifmmode{\tilde{\chi}^0} \else{$\tilde{\chi}^0$} \fi}
\def\sw{\ifmmode{\tilde{\chi}} \else{$\tilde{\chi}$} \fi}

\newcommand{\gsim}{\;\raisebox{-0.9ex}
           {$\textstyle\stackrel{\textstyle >}{\sim}$}\;}
\newcommand{\lsim}{\;\raisebox{-0.9ex}{$\textstyle\stackrel{\textstyle<}
           {\sim}$}\;}

\newcommand{\be}{\begin{eqnarray}}
\newcommand{\ee}{\end{eqnarray}}
\newcommand{\nee}{\nonumber\end{eqnarray}}
\newcommand{\nn}{\nonumber\\}

\newcommand{\bee}[1]{\begin{equation} \label{(#1)}}
\newcommand{\eee}{\end{equation}}
\newcommand{\baq}[1]{\begin{eqnarray} \label{(#1)}}
\newcommand{\eaq}{\end{eqnarray}}
\newcommand{\rf}[1]{(\ref{(#1)})}

\newcommand{\slashed}[1]{\not\!#1}

\begin{document}

\vspace*{-1cm}
\begin{flushright}
  hep-ph/0610234 \\
  UWThPh-2006-26 
\end{flushright}

\vspace*{1.4cm}

\begin{center}

{\Large {\bf CP asymmetries in scalar bottom quark decays 
}}\\

\vspace{2cm}

{\large
A.~Bartl$^a$, E.~Christova$^b$, K.~Hohenwarter-Sodek$^a$, T.~Kernreiter$^a$}

\vspace{2cm}

$^a${\it Institut f\"ur Theoretische Physik, Universit\"at Wien, A-1090
Vienna, Austria}
$^b${\it Institute for Nuclear Research and Nuclear Energy,
Sofia 1784, Bulgaria}

\end{center}

\vspace{1cm}

\begin{abstract}

We propose CP asymmetries based on triple product correlations in the decays
$\sb_m \to t \ti\chi_j^-$ with subsequent decays of $t$ and $\ti\chi_j^-$.
For the subsequent $\ti\chi_j^-$ decay into a leptonic final state
$\ell^- \bar{\nu} \ti\chi^0_1$ we consider the three possible decay chains
$\ti\chi_j^- \to \ell^-\bar{\sn} \to \ell^- \bar{\nu}\nt_1$,
$\ti\chi_j^- \to \sl^-_n\bar{\nu} \to \ell^-
\bar{\nu}\nt_1$ and $\ti\chi_j^- \to W^-\nt_1 \to \ell^- \bar{\nu}\nt_1$. 
We consider two classes of CP asymmetries. In the first class it must be
possible to distinguish between different leptonic $\ti\chi_j^-$ decay chains,
whereas in the second class this is not necessary. We consider also the 
2-body decay $\ti\chi_j^- \to  W^- \tilde\chi^0_1$, and we assume 
that the momentum of the $W$ boson can be measured.
Our framework is the minimal supersymmetric standard model with complex
parameters. The proposed CP asymmetries are non-vanishing due to non-zero
phases for the parameters $\mu$ and/or $A_b$. We present numerical 
results and estimate the observability of these CP asymmetries.

\end{abstract}



\newpage


\section{Introduction}

In the Minimal Supersymmetric Standard Model (MSSM)
\cite{susy,guha} the higgsino mass parameter $\mu$ and several of the
Supersymmetry (SUSY) breaking parameters are complex in general.
Among the SUSY breaking parameters the trilinear scalar couplings $A_{f}$
and two of the gaugino mass parameters $M_1$ and $M_3$ ($M_2$ is usually
chosen to be real by redefining the fields) can be complex.

Current experimental upper bounds on the electric dipole moments (EDM)
impose restrictions on the SUSY parameters that appear in supersymmetric models,
in particular on their phases. To which extent the size of the phases have to be
restricted, however, strongly depends on the underlying model.
For instance, while only relatively small values of the phase of $\mu$, 
$|\phi_\mu|\lsim 0.1$, are allowed in several versions of the MSSM with selectron
masses of the order $100$~GeV \cite{edm},
this restriction may disappear if lepton flavour violating terms
are included \cite{Bartl:2003ju} or if
the masses of the first and second generation scalar fermions are large
(above the TeV scale) while the masses of the third
generation scalar fermions are small (below $1$~TeV) \cite{cohen}.
Recently it has been pointed out that for large trilinear scalar couplings $|A|$
one can simultaneously fulfill the EDM constraints of electron, neutron, and that
of the atoms $^{199}$Hg and $^{205}$Tl, where at the same time,
$\phi_\mu\sim O(1)$ \cite{YaserAyazi:2006zw}.
The restictions on the size of the phases of the trilinear
couplings of the 3rd generation scalar fermions are far less important as
their contributions to the EDMs appear only at two-loop level \cite{Pilaftsis}.

The various CP phases can have a big influence on
the production and decay of supersymmetric particles.
In particular the influence of the phases $\phi_{A_{\tau,t,b}}$ of the trilinear
scalar coupling parameters on various observables
(e.g. scalar fermion masses, cross sections, decay widths) can be important
\cite{Bartl:2002uy,Bartl:2003he}. However, a measurement of solely CP-even observables
cannot be sufficient to unambiguously determine the SUSY parameters.
Moreover, in order to clearly demonstrate that CP is violated, CP-odd observables
have to be measured. Rate asymmetries have been proposed where the influence of the
SUSY CP phases arise due to loop corrections (see for instance \cite{Bernreuther:1995nw}).
Another important class of CP-odd observables are based on
triple product correlations (for an introduction see \cite{Valencia:1994zi}).
They arise already at tree-level and
allow to define various CP asymmetries which are sensitive to the
different CP phases. Such CP asymmetries have been proposed and analyzed
for various SUSY processes (see for instance \cite{Choi:1999cc,Bartl:2004jr}).

Recently, it has been shown~\cite{Bartl:2004jr} that
triple product correlations among the decay products of the
scalar top decay $\st \to t \nt$ followed by the decays of $t$ and $\tilde\chi^0$,
allow us to obtain information on CP violation
in the scalar top system.
Along the same line of the study performed in \cite{Bartl:2004jr}, in the
present paper we analyze triple product correlations that arise in the decays
of the scalar bottoms $\tilde b_m$. We focus on the influence
of CP violation in the scalar bottom system, in particular on the
influence of the phase of the trilinear scalar coupling
parameter $A_b$, $\phi_{A_b}$.

We study the decay
\be
\sb_m \to t\chm_j\label{sb}~,
\ee
followed by the subsequent decays of the top quark $t$ and
the chargino $\chm_j$. We work in the approximation where $t$ and $\tilde\chi^-_j$
are both produced on mass-shell.
As the top quark does not form a bound state this implies that
both $t$ and $\tilde\chi^-_j$ decay with definite momenta and polarizations.
Their polarizations can be retrieved from the distributions of their decay products.
We consider the decays of the top quark
\be
t\to b~W^+~\qquad {\rm and}\qquad t\to b~l^+ \nu_l~(b~c~\bar s)~,\label{t}
\ee
and the following three possible decay chains for
$\chm_j$:
\begin{eqnarray}
\chm_j  \to &\ell_1^-\bar{\sn}~\quad \to &\ell_1^-\bar{\nu}\nt_1~,\label{I}\\
\chm_j  \to &\sl_n^-\bar{\nu}~\quad  \to  &\ell_2^-\bar{\nu}\nt_1~,\label{II}\\
\chm_j  \to & W^- \nt_1~  \to  &\ell_3^-\bar{\nu} \nt_1~,\label{III}
\end{eqnarray}
which lead to the final states
\be
\chm_j \to \ell^-\bar{\nu} \nt_1~,\qquad \ell=e,\mu,\tau~.
\label{3body}
\ee
We shall consider each of the decays (\ref{I}),(\ref{II}),(\ref{III}) separately. 
The subscript of the
leptons, $\ell_1$, $\ell_2$, $\ell_3$, is used in order to distinguish them in the
different decay chains. For simplicity we shall work in the narrow width approximation
for the intermediate particles in (\ref{I})--(\ref{III}), i.e.
we assume that these particles are produced on-mass-shell.

We consider also the 2-body decay of $\chm_j$:
\be
\chm_j \to W^- \nt_1~,\label{2body}
\ee
assuming the momentum of the final $W$ boson can be reconstructed,
which is possible for hadronic decays.

We consider the triple products 
\be
\mathcal O = {\bf q}_1 \cdot ({\bf q}_2 \times {\bf q}_3)\equiv
({\bf q}_1 {\bf q}_2 {\bf q}_3)~,
\label{tripleprod}
\ee
where ${\bf q}_i$ are any 3-vectors of the particles in the considered process.
With the help of the triple products ${\mathcal O}$, Eq.~(\ref{tripleprod}), we 
define the T-odd observables (up-down asymmetries):
\bee{eq:cpasy}
A_T\equiv\frac{\int d\Omega ~sgn({\mathcal O}) ~d\Gamma/d\Omega}
{\int d\Omega ~d\Gamma/d\Omega} =
\frac{N[{\mathcal O}> 0]-N[{\mathcal O} < 0]}
{N[{\mathcal O}> 0]+N[{\mathcal O} < 0]}~,
\eee
where $d\Gamma$ stands for the differential decay width and
$d\Omega$ involves the angles of integration.
The left hand side of Eq.~\rf{eq:cpasy} shows how the
asymmetries are calculated, whereas the right hand side
exemplifies how they are measured in experiment:
$N[{\mathcal O} > (<)~ 0]$ is the number
of events for which ${\mathcal O}> (<) ~0$.

The paper is organized as follows.
In section \ref{sec:2} we present the results of our
calculations in compact form using the formalism of \cite{K&T}.
We propose several T-odd asymmetries in section \ref{sec:3} and
point out how the corresponding CP asymmetries can be obtained.
In section \ref{sec:4} we perform a numerical analysis of the
CP asymmetries proposed and estimate their observability.
Finally, we summarize in section \ref{concl}.

\section{Formalism\label{sec:2}}
%
In order to obtain analytic expressions for the sequential processes
(\ref{sb})--(\ref{2body}) we shall use
the formalism of Kawasaki, Shirafuji and Tsai \cite{K&T}. According to that
formalism the differential decay rates of (\ref{sb})--(\ref{2body}),
when spin-spin correlations are taken into account, can be written as
\bee{eq:gen}
d\Gamma=d\Gamma(\ti b_m \to t \ti\chi^-_j)\,\frac{E_t}{m_t\Gamma_t}\,
d\Gamma (t\to ...)\,\frac{E_{\chi_j}}{m_{\chi_j} \Gamma_{\chi_j}}
\,d\Gamma(\tilde\chi^-_j\to ...)~,
\eee
where $d\Gamma (t \to ...)$
and $d\Gamma (\tilde\chi^-_j\to ...)$ are the
differential decay rates of the unpolarized top and unpolarized chargino.
The factors $E_{\chi_j}/(m_{\chi_j}{\Gamma}_{\chi_j})$
and $E_t/(m_t\Gamma_t)$ stem from the narrow width approximation used for $t$ and
$\tilde\chi^-_j$, $\Gamma_t$ and $\Gamma_{\chi_j}$ are the total widths of $t$ and
$\tilde\chi^-_j$, and $(E_t, m_t)$ and $(E_{\chi_j}, m_{\chi_j})$ 
are their energies and masses, respectively.
$d\Gamma(\ti b_m \to t \ti\chi^-_j)$ is the differential decay rate of the 
scalar bottom $\sb_m$ into a top
quark with the polarization 4-vector $\xi_t^\alpha$
and a chargino with the polarization 4-vector $\xi_{\chi_j}^\alpha$.

In the scalar bottom rest frame, we have: 
\bee{eq:gammast}
d\Gamma(\ti b_m \to t~\tilde\chi^-_j)
=\frac{2}{ m_{\ti b_m}}\vert A\vert^2\,d\Phi_{\tilde b_m}~,
\eee
where $m_{\ti b_m}$ is the mass of the decaying scalar bottom and
the phase space element $\Phi_{\sb}$ is given in Eq.~\rf{eq:phisbott} 
in Appendix \ref{app:phasespace}. For the matrix element $A$ we have
\be
A=g\bar u(p_t) (k_{mj}^{\tilde b}P_L + l_{mj}^{\tilde b}P_R)v(p_{\chi_j})~,
\ee
where $P_{L,R}=\frac{1}{2}(1\mp \gamma_5)$,
$g$ is the $\mathrm{SU(2)}$ gauge coupling constant
and the couplings are given in Eq.~\rf{eq:couplsbott} in Appendix \ref{app:lagrange}.
For the evaluation of $|A|^2$ we use the spin density matrices of $t$
and $\tilde\chi^-_j$:
\be
\rho (p_t) = \Lambda (p_t)\,\frac{1+\gamma_5\slashed{\xi}_t}{2}~,\qquad
\rho (-p_{\chi_j}) = - \Lambda (-p_{\chi_j})\,\frac{1+
\gamma_5\slashed{\xi}_{\chi_j}}{2}~,
\ee
with
\be
\Lambda (p_t) = \slashed{p}_t +m_t~,\qquad \Lambda (p_{\chi_k}) =
\slashed{p}_{\chi_j}+m_{\chi_j}~.
\ee
The matrix element squared is then given by
\begin{eqnarray}
\vert A\vert^2&=& \frac{g^2}{2}
\left\{(\vert l_{mj}^{\ti b}\vert^2 + \vert k_{mj}^{\ti b}\vert^2)\,
[(p_{\chi_j} p_t) +
m_{\chi_j} m_t \, (\xi_{\chi_j}\xi_t)]\right.\nonumber\\
&&\qquad -(\vert l_{mj}^{\ti b}\vert^2 - \vert k_{mj}^{\ti b}\vert^2)\,
[m_t (p_{\chi_j}\xi_t)+m_{\chi_j}\,(\xi_{\chi_j} p_t)]\nonumber\\
&&\qquad -\,2\,{\Re e}(l_{mj}^{\ti b*}k_{mj}^{\ti b})\,
[m_{\chi_j} m_t -(p_{\chi_j}\xi_t)(\xi_{\chi_j} p_t)+
(p_{\chi_j} p_t)(\xi_{\chi_j} \xi_t)] \nonumber\\
&&\qquad +\left.2\, {\Im m}(l_{mj}^{\ti b*}k_{mj}^{\ti b})~
\varepsilon^{\a\b\gamma\delta}~p_{\chi_j\a}~\xi_{\chi_j\b}~
\xi_{t\gamma}~p_{t\delta}\right\}
\label{A2}
\end{eqnarray}
where we use the convention $\varepsilon^{0123}=1$.
The polarization 4-vector $\xi_t$ is determined through the top quark
decays (\ref{t}) and the polarization 4-vector
$\xi_{\chi_j}$ is determined through the $\ti\chi_j$ decays
(\ref{I})--(\ref{2body}). In the following we calculate the
polarization 4-vectors $\xi_t$ and $\xi_{\chi_j}$ as well as the
differential decay rates of $t$ and $\chm_j$ for their various decays (\ref{t}) and
(\ref{I})--(\ref{2body}).
Some of the calculations are quite analogous to those carried out in \cite{Bartl:2004jr}
and in these cases we present the results only.

\subsection{Decay rates for $\ti\chi^-_j\to\ell_1^-\bar{\sn}\to\ell_1^- \bar{\nu}\ti\chi^0_1$}
%

The polarization vector of the top quark was obtained in \cite{Bartl:2004jr} and
here we present the results for completeness. The polarization 4-vector of the
top quark determined through the decay $t\to b~W^+$, that we shall denote by
$\xi_b$, equals
\be
\xi_b^\alpha =\alpha_b\,\frac{m_t}{(p_tp_b)}\,
\left(p_b^\alpha -\frac{(p_tp_b)}{m_t^2}\,p_t^\alpha\right)~,
\qquad \alpha_b= \frac{m_t^2-2m_W^2}{m_t^2+ 2m_W^2}~.
\label{xitopb}
\ee
For the polarization vector of the top quark determined in $t\to b~W^+\to b~l^+ \nu$
(and equivalently for $t\to b~W^+\to b~c~\bar{s}$, where we substitute the
the $c$ quark for the lepton),
that we denote by $\xi_l$, we have
\be
\xi_l^\alpha
=\alpha_l\,\frac{m_t}{(p_tp_l)}\,
\left(p_l^\alpha -\frac{(p_tp_l)}{m_t^2}\,p_t^\alpha\right)~,
\qquad
\alpha_l=-1~.\label{xitopl}
\ee

The polarization vector of $\chm_j$ is determined solely through the decay
$\chm_j\to \ell_1^-\bar{\sn}$, as the subsequent decay of $\sn$, being a scalar particle,
does not contribute. We obtain:
\be
\xi_{\chi_j}^\alpha
=\alpha_{\sn}\frac{m_{\chi_j}}{(p_{\chi_j} p_{\ell_1})}~
\left(p_{\ell_1}^\alpha -
\frac{(p_{\chi_j} p_{\ell_1})}{m^2_{\chi_j}}~p_{\chi_j}^\a\right)~,
\qquad
\alpha_{\ti\nu}=\frac{|l_j^{\ti\nu}|^2-|k_j^{\ti\nu}|^2}
{|l_j^{\ti\nu}|^2+|k_j^{\ti\nu}|^2}~.
\label{xi1}
\ee

Further, according to Eq.~\rf{eq:gen},
we need the differential decay rates of $t$ and $\ti\chi_j^-$.
The distribution of the leptons in the sequential decay (\ref{I}),
in the narrow width approximation for $\sn$, is given by
\bee{eq:gammachar1}
d\Gamma^{\rm I}_{\chi_j} (\tilde\chi^-_j\to \ell_1^- \bar{\nu} \ti\chi^0_1)=
d\Gamma (\ti\chi^-_j\to\ti \ell_1^- \bar{\sn})~BR(\bar{\sn} \to \bar{\nu}\nt_1)~,
\eee
where $BR(\bar{\sn} \to \bar{\nu}\nt_1)$ is the branching ratio of the decay
$\bar{\sn} \to \bar{\nu}\nt_1$ and
\bee{eq:def}
d\Gamma (\ti\chi^-_j\to \ell_1^-\bar{\ti\nu})=\frac{g^2\,(|k^{\ti\nu}_j|^2+
|l^{\ti\nu}_j|^2)\,(p_{\chi_j} p_{\ell_1})}{2 E_{\chi_j}}~d\Phi^1_{\chi_j}~,
\eee
where the couplings are given in Eq.~\rf{eq:snucoupl} in Appendix \ref{app:lagrange}
and the phase space element $d\Phi^1_{\chi_j}$ is given in Eq.~(\ref{eq:phi1}) in
Appendix \ref{app:phasespace}.
The differential decay rates of the top quark are (see for instance \cite{Bartl:2004jr}):
\bee{eq:gammab}
d\Gamma (t \to b W^+)=  \frac{g^2(m_t^2-m_W^2) \,
(2m_W^2+m_t^2)}{8E_t\,m_W^2}~d\Phi_t^b~,
\eee
\bee{eq:gammal}
d\Gamma (t \to bl^+\nu)=\frac{g^4\,\pi (p_tp_l)
\,(m_t^2-2(p_tp_l))}{2E_t\, m_W\,\Gamma_W} \,d\Phi_t^l~,
\eee
with $d\Phi_t^{b,l}$ given in Eqs.~(\ref{eq:pstopW}) and (\ref{eq:pstopl})
in Appendix \ref{app:phasespace}.

The angular distributions of the decay products of $t$ and
$\ti\chi_j^-$ decay mode (\ref{I}) are obtained by inserting the differential decay rate
of the scalar bottom, Eq.~\rf{eq:gammast}, the differential decay rates of the
top quark, Eqs.~\rf{eq:gammab} and \rf{eq:gammal}, and the
differential decay rate of the chargino, Eq.~\rf{eq:gammachar1},
into Eq.~\rf{eq:gen}, where we use the appropriate polarization vectors
as given in Eqs.~(\ref{xitopb})--(\ref{xi1}).
The differential decay rates of $\ti b_m$ then read
\be
d\Gamma_f^{\rm I} &=&N_f\,\frac{g^6\,BR (\sn \to \nu \ti\chi_1^0)\,
(p_{\chi_j}p_{\ell_1})~\left(|l_j^{\sn}|^2+|k_j^{\sn}|^2\right)}
{8~m_{\sb}~m_t\Gamma_t\,m_{\chi_j}\Gamma_{\chi_j}}\nn
&&\times
\biggl\{(|l_{mj}^{\ti b}|^2+|k_{mj}^{\ti b}|^2)~
(p_{\chi_j} p_t)-2~{\Re e}
(l_{mj}^{\ti b*}k_{mj}^{\ti b})\,m_{\chi_j} m_t+\cdots\biggr.\nn
&&
\biggl.+2~{\Im m}(l_{mj}^{\ti b*}k_{mj}^{\ti b})~
\alpha_f~\alpha_{\sn} \frac{m_t}{(p_tp_f)}\frac
{m_{\chi_j}}{(p_{\chi_j}p_{\ell_1})}\,m_{\sb}~
({\bf p}_{\ell_1}{\bf p}_f {\bf p}_t)
\biggr\}\,d\Phi_f^{\rm I}~,\label{B1}
\ee
where the subindex $f=b,l$ is to distinguish the two $t$ quark decays in (\ref{t}).
The prefactors in Eq.~(\ref{B1}) are
\be
N_b&=& \frac{(m_t^2-m_W^2)(2m_W^2+m_t^2)}{2\,m_W^2}~,\nn
N_l&=& \frac{g^2\,2\,\pi\,(p_tp_l)(m_t^2-2(p_tp_l))}{m_W\Gamma_W}~,
\ee
and the phase space elements equal
\be
d\Phi_f^{\rm I}= d\Phi_{\sb_m}\, d\Phi_t^f \,d\Phi_{\chi_j}^{1}~.
\ee
In Eq.~(\ref{B1}) we have only included those terms which are needed for
the calculation of the up-down asymmetries in Eq.~\rf{eq:cpasy}.
The omitted terms, represented by dots, are T-even and thus, cannot contribute 
to the numerator of Eq.~\rf{eq:cpasy}. Further, as they depend on the
polarizations of either the top quark or the chargino, they cannot 
contribute to the denominator of Eq.~\rf{eq:cpasy}.

\subsection{Decay rates for $\ti\chi^-_j\to\sl^-_n \bar{\nu} \to \ell_2^- \bar{\nu} \nt_1$}

In order to obtain the angular correlations among the $t$ decay
products and the lepton $\ell_2$ stemming from the $\chm_j$ decay (\ref{II}), we need the
polarization 4-vector of $\chm_j$ determined in the decay (\ref{II}).
As $\sl_n$ is a scalar particle, $\xi_{\chi_j}$ is determined solely in the decay
$\chm_j \to \sl_n^- \bar{\nu}$. We obtain:
\be
\xi_{\chi_j}^{\alpha} =\alpha_{\sl}~\frac{ m_{\chi_j}}
{(p_{\chi_j}p_{\nu})}~\left(p_\nu^\a-
\frac{(p_{\chi_j}p_{\nu})}{m_{\chi_j}^2}\,p_{\chi_j}^\a\right)~,\qquad
\alpha_{\ti\ell}=-1~. \label{xi2}
\ee
The differential decay rate of the decay chain (\ref{II}), in the narrow width
approximation for $\sl_n^-$, reads
\bee{eq:gamma2}
d\Gamma^{\rm II}_{\chi_j}(\tilde\chi^-_j\to \ell_2^-\,\bar{\nu}~\ti\chi^0_1)=
d\Gamma(\tilde\chi^-_j\to \sl^-_n \bar{\nu})\,\frac{E_{\sl}}{m_{\sl}\,\Gamma_{\sl}}\,
d\Gamma(\sl^-_n\to \nt_1 \ell_2^-)~,
\eee
with the differential decay rates for $\tilde\chi^-_j\to \sl^-_n\bar{\nu}$ and
$\sl_n\to \nt_1\ell_2^-$ given by
\be
d\Gamma(\tilde\chi^-_j\to \sl_n^-\bar\nu)&=&\frac{g^2}{2\, E_{\chi_j}}\,
\vert l_{nj}^{\sl}\vert^2 \, (p_{\chi_j}p_\nu )\,d\Phi^{2}_{\chi_j}~,
\label{gammachi2}
\ee
and
\be
d\Gamma (\sl^-_n\to \nt_1\ell_2^-) &=& \frac{g^2}{E_{\sl}}\,
(|a^{\ti\ell}_{nk}|^2+|b^{\ti\ell}_{nk}|^2)\,(p_{\ti\ell}p_{\ell_2})d\Phi_{\sl}~,
\label{gammasl}
\ee
where the couplings are given in Appendix \ref{app:phasespace}
in Eqs.~\rf{eq:albl} and \rf{eq:ll}.
The phase space elements $d\Phi^{2}_{\chi_j}$ and $d\Phi_{\sl}$ are given in
Appendix \ref{app:lagrange} in Eqs.~(\ref{eq:phasespchiW2})
and (\ref{eq:phasespsl}), respectively.

The angular distributions of the decay products of $t$ are
the same as in the previous case.
On the other hand, the angular distribution of the decay products of the
$\ti\chi_j^-$ decay mode (\ref{II}) is given by Eq.~\rf{eq:gamma2} which
we insert into Eq.~\rf{eq:gen} in order to obtain the differential decay rates
of the combined process (\ref{sb}), (\ref{t}) and (\ref{II}).
The polarization vector of the chargino is determined through the
decay (\ref{II}) and is given in Eq.~(\ref{xi2}).
Then the differential decay rates of $\ti b_m$ read
\be
d\Gamma_f^{\rm II}&=&N_f\,
\frac{g^8\,(p_{\chi_j}p_\nu)(p_{\chi^0_1}p_{\ell_2})|l_{nj}^{\sl}|^2
(|a^{\sl}_{nk}|^2+|b^{\sl}_{nk}|^2)}{8 m_{\sb} m_t\Gamma_t
m_{\chi_j}\Gamma_{\chi_j}m_{\sl}\,\Gamma_{\sl}}\nn
&&\times
\biggl\{(|l_{mj}^{\ti b}|^2+|k_{mj}^{\ti b}|^2)
(p_{\chi_j} p_t)-2{\Re e}(l_{mj}^{\ti b*}k_{mj}^{\ti b})\,m_{\chi_j} m_t+\cdots\biggr.\nn
&&+\biggl. 2{\Im m}(l_{mj}^{\ti b*}k_{mj}^{\ti b})
\alpha_f \alpha_{\sl} \frac{m_t}{(p_tp_f)}\frac
{m_{\chi_j}}{(p_{\chi_j}p_\nu)} m_{\sb} ({\bf p}_{\ell_2}{\bf p}_f {\bf p}_t)
\biggr\}\,d\Phi_f^{\rm II}~,\label{B2}
\ee
where the phase space elements equal
\be
d\Phi_f^{\rm II}= d\Phi_{\sb_m}\, d\Phi_t^f \,d\Phi_{\chi_j}^{2}\,d\Phi_{\sl}~.
\ee
As in the previous case, we have omitted those terms in Eq.~(\ref{B2})
(denoted by dots) which are unessential
for the calculation of the up-down asymmetries, Eq.~\rf{eq:cpasy}.

\subsection{Decay rates for $\ti\chi_j^-\to  W^- \ti\chi^0_1 \to l_3^-\bar\nu\ti\chi^0_1$}
%

When the decay of $\ti\chi^-_j$  proceeds via the $W^-$ boson exchange, (\ref{III}),
the polarization 4-vector $\xi_{\chi_j}$ is parameterized
by two components that are in the $\chm_j$ decay plane and
a component normal to it. It can be written completely general as
\be
\xi_{\chi_j}^{\alpha} = P_\ell Q_{\ell}^{\alpha}+P_\nu Q_\nu^{\alpha}
+D^{CP}\varepsilon^{\a\b\gamma\delta}~p_{\ell_3\b}~p_{\nu\gamma}~p_{\chi_j\delta}
\label{xiW}
\ee
where the 4-vectors $Q^\alpha_\ell$ and $Q^\alpha_\nu$ are in
the decay plane of $\ti\chi^-_j$:
\be
&& Q^\alpha_\ell=p_{\ell_3}^\alpha-
\frac{(p_{\ell_3} ~p_{\chi_j})}{m^2_{\chi_j}}p^\alpha_{\chi_j}~, \qquad
Q^\alpha_\nu=p_\nu^\a-\frac{(p_\nu ~p_{\chi_j})}{m^2_{\chi_j}}p^\alpha_{\chi_j}~,
\ee
and $\varepsilon^{\a\b\gamma\delta}p_{\ell_3\b}p_{\nu\gamma} p_{\chi_j\delta}$
is orthogonal to it. For the components in the decay plane we obtain
\be
P_\ell &=&\frac{m_{\chi_j}|O^L_{1j}|^2\left(2(p_\nu p_{\chi_j})-m^2_W\right)
-2 m_{\chi^0_1}(p_\nu p_{\chi_j})
\Re e (O_{1j}^{L*}O_{1j}^R)}{|C|^2}~,\nonumber \\
P_\nu &=&\frac{
-m_{\chi_j}|O^R_{1j}|^2\left(2(p_{\ell_3} p_{\chi_j})-m^2_W\right)+2 m_{\chi^0_1}
(p_{\ell_3} p_{\chi_j}) \Re e (O_{1j}^{L*}O_{1j}^R)}
{|C|^2}~,\label{Ql}
\ee
with
\be
|C|^2&=&-m_W^2\left[|O^L_{1j}|^2 (p_{\ell_3} p_{\chi_j})+ |O^R_{1j}|^2
(p_\nu p_{\chi_j})+
 m_{\chi^0_1} m_{\chi_j} \Re e(O_{1j}^{L*}O_{1j}^R)\right]\nn
&&+ 2(p_{\ell_3} p_{\chi_j})(p_\nu p_{\chi_j})(|O^L_{1j}|^2+|O^R_{1j}|^2)~,
\ee
where the couplings are given in Appendix \ref{app:lagrange} in Eq.~\rf{eq:OLOR}.
The component normal to the decay plane reads
\be
D^{CP}=\frac{2 m_{\chi^0_1} \Im m({O^L_{1j}}^*O^R_{1j})}{\vert C\vert^2}~.\label{D}
\ee
The component $D^{CP}$ is sensitive to CP violation in the
$\ti\chi_j^-\ti\chi_1^0W^+$ couplings, i.e.
to the phases $\phi_\mu$ and $\phi_{M_1}$.
The decay rate distribution of
$\ti\chi_j^-\to  W^- \ti\chi^0_1\to \ell_3^-\bar\nu\ti\chi^0_k$ is given by
\be
d\Gamma_{\chi_j}^{\rm III}(\tilde\chi^-_j\to \ell_3^-\bar\nu~\ti\chi^0_1)=
\sum_\pm\frac{g^4\,\pi }{m_W\,\Gamma_W\,E_{\chi_j}\,}~|C|^2\,
d\Phi^{\rm III}_{\chi_j}~,\label{gammachi3}
\ee
where
$d\Phi^{\rm III}_{\chi_j}=\frac{1}{2 \pi}(d\Phi^3_{\chi_j})^\pm~d\Phi_W^3$ with
$(d\Phi^3_{\chi_j})^\pm$ being the phase space element for the decay
$\ti\chi_j^- \to W^-\ti\chi_1^0$, Eq.~\rf{eq:PchjW} in Appendix \ref{app:phasespace},
and $d\Phi_W^3$ is the phase space element for the decay
$W^-\to\ell^-_3\bar\nu$, Eq.~(\ref{PhW3}).

The angular distributions of the decay products of $\ti b_m$,
where the chargino decays according to (\ref{III}), can now be obtained in
the same manner as in the previous two cases.
Again we only quote the terms that are essential
for the calculation of the up-down asymmetries in Eq.~\rf{eq:cpasy}:
\be
d\Gamma_f^{\rm III} &=& \sum_\pm N_f~\frac{g^8~\pi~|C|^2}
{4 m_{\sb }m_t\Gamma_tm_{\chi_j}\Gamma_{\chi_j}~m_W\Gamma_W}\,\nn
&&\times\biggl\{(|l_{mj}^{\ti b}|^2+|k_{mj}^{\ti b}|^2)~
(p_{\chi_j} p_t)-2{\Re e}(l_{mj}^{\ti b*}k_{mj}^{\ti b})m_{\chi_j}m_t+\cdots\biggr.\nn
&&\biggl. +2\,\alpha_f\, {\Im m}(l_{mj}^{\ti b*}k_{mj}^{\ti b})\,
\frac{m_t}{(p_tp_f)}(P_\ell-P_\nu) m_{\sb}({\bf p}_{\ell_3}{\bf p}_f {\bf p}_t)]
\biggr\}~d\Phi_f^{\rm III}~,
\label{A3}
\ee
with
\be
d\Phi_f^{\rm III} = d\Phi_{\sb}~ d\Phi_t^f~ d\Phi_{\chi_j}^{\rm III}~,
\ee
where the sum in Eqs.~(\ref{gammachi3}) and (\ref{A3}) corresponds to the two kinematical
solutions for $E_{\ell_3}$ (for details see Appendix \ref{app:phasespace}).

In principle, the normal component of the chargino polarization
vector in Eq.~(\ref{xiW}) will also give rise to triple products
proportional to $\Im m({O^L_{kj}}^*O^R_{kj})$.
However, in order to be sensitive to these triple products, the
reconstruction of the decay plane of the
chargino would be necessary. In practice, this cannot be
accomplished, because the neutrino as well as the neutralino
escape detection in experiment.

\subsection{Decay rates for $\ti\chi^-_j \to W^-\ti\chi^0_1$}

Finally we consider the two-body decay mode of $\ti\chi^-_j$ (\ref{2body}).
The polarization 4-vector of $\ti\chi^-_j$ in this case is given as
\be
\xi_{\chi_j}^\alpha =\a_W \frac{m_{\chi_j}}{(p_{\chi_j}p_W)}
\left(
p^\alpha_W-\frac{(p_W ~p_{\chi_j})}{m^2_{\chi_j}}p^\alpha_{\chi_j}
\right)~,
\ee
with
\be
\a_W=2\left(\frac{|O_{1j}^L|^2-|O_{1j}^R|^2}
{|C_W|^2}\right)
\left(\frac{m^2_{\chi_j}-2 m_W^2 -m^2_{\chi^0_1}}{m_W^2}\right)
(p_{\chi_j}p_W)~,
\ee
where
\baq{eq:CW}
|C_W|^2&=&(|O^L_{1j}|^2+|O^R_{1j}|^2)
\left[ \frac{(m^2_{\chi^0_1}+m^2_{\chi_j})m^2_W+
(m^2_{\chi^0_1}-m^2_{\chi_j})^2-2 m^4_W}{m^2_W}
\right] \nonumber \\[3mm]
& & {} - 12\,{\Re}e\,({O^L_{1j}}^*O^R_{1j})\,m_{\chi^0_1} m_{\chi_j}~.
\eaq
The decay rate distribution of
$\ti\chi_j^-\to W^- \ti\chi^0_1$ is given by
\be
d\Gamma_{\chi_j}^{W}(\tilde\chi^-_j\to W^-~\ti\chi^0_1)=
\sum_\pm \frac{g^2}{4 E_{\chi_j}}|C_W|^2(d\Phi^3_{\chi_j})^\pm~.\label{decayratetwobody}
\ee

The angular distribution of the decay products of $\ti b_m$,
with the chargino two-body decay (\ref{2body}), is given by
\be
d\Gamma_f^{W} &=&\sum_\pm N_f\,\frac{g^6~|C_W|^2}
{16~m_{\sb}~m_t\Gamma_t\,m_{\chi_j}\Gamma_{\chi_j}}\nn
&&\times
\biggl\{(|l_{mj}^{\ti b}|^2+|k_{mj}^{\ti b}|^2)~
(p_{\chi_j} p_t)-2~{\Re e}
(l_{mj}^{\ti b*}k_{mj}^{\ti b})\,m_{\chi_j} m_t+\cdots\biggr.\nn
&&
\biggl.+2~{\Im m}(l_{mj}^{\ti b*}k_{mj}^{\ti b})~
\alpha_f~\alpha_{W} \frac{m_t}{(p_tp_f)}\frac
{m_{\chi_j}}{(p_{\chi_j}p_{\ell_1})}\,m_{\sb}~
({\bf p}_{W}{\bf p}_f {\bf p}_t)
\biggr\}~d\Phi_f^{W}~,
\label{distwobody}
\ee
with
\be
d\Phi_f^{W} = d\Phi_{\sb}~ d\Phi_t^f~(d\Phi^3_{\chi_j})^\pm~,
\ee
where again we have quoted only the terms that contribute
to the up-down asymmetries in Eq.~\rf{eq:cpasy}.
The sum in Eq.~(\ref{distwobody}) is due to the two kinematical solutions for
$|{\bf p}_W|$ (for details see Appendix \ref{app:phasespace}).

\section{T-odd asymmetries \label{sec:3}}

A general definition of the T-odd observables which we study
in this paper has been given in Eq.~\rf{eq:cpasy}.
For the following it is convenient to introduce a shorthand
notation for the various T-odd asymmetries to be studied below:
\be
A_{ijk}=\frac{ N\left[({\bf p}_{i} {\bf p}_j {\bf
p}_{k})
> 0 \right] -N\left[({\bf p}_{i}{\bf p}_j {\bf
p}_{k}) < 0
\right]} {N\left[({\bf p}_{i}{\bf p}_k {\bf
p}_{k}) > 0 \right]
+N\left[({\bf p}_{i}{\bf p}_j {\bf
p}_{k}) < 0 \right]}~,
 \label{Asym}
\ee
where $N\left[({\bf p}_{i} {\bf p}_j {\bf p}_{k})> 0 \right]$ 
($N\left[({\bf p}_{i} {\bf p}_j {\bf p}_{k})< 0 \right]$) are the number
of events with $({\bf p}_{i} {\bf p}_j {\bf p}_{k})> 0$
($({\bf p}_{i} {\bf p}_j {\bf p}_{k})< 0$).
The indices $i,j,k$ specify the observed particles appearing in the considered
decay mode of $\ti b_m$. We choose the convention that
${\bf p}_{i}$ denotes the momentum of a particle originating from the 
$\ti\chi_j^-$ decay, ${\bf p}_{j}$ denotes the momentum of a fermion from the 
top quark decay and ${\bf p}_{k}$ either denotes the momentum of the top quark 
itself or of another particle stemming from the decay of the top quark. 
According to the different
decay channels we group  the considered  triple products as follows:

\noi
${\bf(I)}$ If the 3-body decay $\ti\chi^-_j \to \ell^-_i \bar{\nu}\ti\chi^0_1$
is considered, the only detectable particles are
the final charged leptons $\ell_1^-,\ell_1^-,\ell_3^-$. We shall
distinguish two classes of asymmetries depending on whether 
the leptons $\ell_1^-,\ell_2^-,\ell_3^-$,
originating from the different decay chains (\ref{I})--(\ref{III}),
are distinguishable or not.

{\bf 1.} First we define the T-odd asymmetries where the leptons from the
$\ti\chi_j^-$ decays (\ref{I})--(\ref{III}) can be distinguished
\footnote{In principle, the leptons from the decays (\ref{I})--(\ref{III}) can be
distinguished through their different angular or energy distributions.}.
The triple products on which the T-odd asymmetries are based in this case, are given by
\be
&&({\bf p}_{\ell^-_i} {\bf p}_b {\bf
p}_{t}) \quad
{\rm and } \quad ({\bf p}_{\ell^-_i} {\bf p}_b {\bf p}_{W^+})
~,\quad
{\rm when}\quad t\to b W^+\to bq \bar{q}',
 \label{Asymbi}\\
&& ({\bf p}_{\ell^-_i} {\bf p}_{l^+} {\bf
p}_{b}),\quad
{\rm  when}\quad t\to b W^+\to bl^+\nu,
 \label{Asymli}\\
&&({\bf p}_{\ell^-_i} {\bf p}_{c} {\bf
p}_{t}),\quad ({\bf p}_{\ell^-_i} {\bf p}_{c} {\bf
p}_{b})\quad {\rm and } \quad ({\bf p}_{\ell^-_i} {\bf p}_{c} {\bf p}_{s}),\quad
{\rm  when}\quad t\to b W^+\to b c\bar{s},
\label{Asymci}
\ee
where for the triple products in (\ref{Asymci})
it is necessary to identify the $c$ quark which is expected to
be possible with reasonable efficiency and purity 
\cite{Damerell,flavortagging1,flavortagging2}.
With the triple products in (\ref{Asymbi})--(\ref{Asymci}) the associated
T-odd asymmetries can be defined according to Eq.~(\ref{Asym}), where
in the following we use the notation $A_{\ell^-_ibt}$ and $A_{\ell^-_ibW^+}$ for
the T-odd asymmetries based on the triple products in (\ref{Asymbi}) etc.
Note that $A_{\ell^-_ibt}$ and $A_{\ell^-_ibW^+}$ have the same
value due to momentum conservation.

{\bf 2.}
We define a second class of T-odd asymmetries where it is not
necessary to distinguish the different leptonic $\ti\chi^-_j$ decay chains,
Eqs.~(\ref{I})--(\ref{III}). This class of T-odd asymmetries is based on the
triple products as given in (\ref{Asymbi})--(\ref{Asymci})
where $\ell^-_i$ is replaced by $\ell^-$. Then
$N\left[({\bf p}_{\ell^-}{\bf p}_b {\bf p}_t) > 0 \right] $ in Eq.~(\ref{Asym})
means
\be
\!\!\!\!\!\!N\left[({\bf p}_{\ell^-}{\bf p}_b {\bf p}_t) > 0 \right] =
N\left[({\bf p}_{\ell_1}{\bf p}_b {\bf p}_t) > 0 \right] +
N\left[({\bf p}_{\ell_2}{\bf p}_b {\bf p}_t) > 0 \right] +
N\left[({\bf p}_{\ell_3}{\bf p}_b {\bf p}_t) > 0 \right].\nonumber
\ee
For this class of T-odd asymmetries we will use the
notation $A_{\ell^-bt}$ etc. The following formula relates
$A_{\ell^-jk}$ to the asymmetries $A_{\ell^-_ijk}$
and the branching ratios
$BR_{\ell_i}\equiv BR(\ti\chi^-_j\to \ell^-_i\bar\nu\ti\chi^0_1)$
of the decay chains (\ref{I})--(\ref{III}):
\be
A_{\ell^-jk}= \frac{BR_{\ell^-_1}}{BR_{\ell^-}}A_{\ell^-_1jk} +
\frac{BR_{\ell^-_2}}{BR_{\ell^-}}A_{\ell^-_2jk} +
 \frac{BR_{\ell^-_3}}{BR_{\ell^-}}A_{\ell^-_3jk}~,
\label{sumasy}
\ee
where we have introduced the shorthand notation
$BR_{\ell^-}$:
\be
BR_{\ell^-} = BR_{\ell^-_1}+BR_{\ell^-_2}+BR_{\ell^-_3}~.
\ee
This formula allows us to calculate the contribution of
$A_{\ell^-_ijk}$ to the  asymmetry $A_{\ell^-jk}$, depending on the branching
ratios of the different decay modes
of $\ti\chi^-_j$.

\noi
${\bf(II)}$
If we consider the 2-body decay mode $\ti\chi^-_j \to W^- \ti\chi^0_1$,
where the $W$ boson decays hadronically so that its momentum vector
can be reconstructed, we can define analogous triple products
as in (\ref{Asymbi})--(\ref{Asymci}) with $\ell^-_i$ replaced by $W^-$.
For the corresponding T-odd asymmetries again we use the notation
$A_{W^-bt}$ etc.

At the end of this section, we discuss how CP-odd asymmetries can be obtained from the
T-odd asymmetries defined above. It is well known that a non-zero value of the
considered T-odd asymmetries does not necessarily imply that the CP symmetry is violated
since final state interactions give rise (although only at loop level) to the same
asymmetries. In order to identify a genuine signal of CP violation one needs to consider
also the C-conjugate decay. T-odd asymmetries that are based on triple products analogous
to the one given in (\ref{Asymbi})--(\ref{Asymci}) can be defined for the charge
conjugate decay ${\bar{\tilde b}}_m\to\ti\chi^+_j {\bar t}$ as well, and we denote them
by ${\overline A}_{i j k}$. One finds that the term of the matrix element
squared for the C-conjugate decay ${\bar{\tilde b}}_m\to\ti\chi^+_j {\bar t}$ that
comprises the triple product has the same sign as the corresponding term for the decay
$\sb_m\to\ti\chi^-_j t$. Thus, true CP violating asymmetries are obtained when summing
the T-odd asymmetries that arise in the decays $\sb_m\to\ti\chi^-_j t$ and ${\bar{\tilde
b}}_m\to\ti\chi^+_j {\bar t}$:
\be
A_{ijk}^{\rm CP}=\frac{A_{ijk}+{\overline A}_{ ij k}}{2}~.\label{AssymCP}
\ee

\section{Numerical results \label{sec:4}}

Now we study numerically the CP asymmetries defined in the previous section, where 
we focus on their dependence on the CP phases, in particular on $\phi_{A_b}$.
All CP asymmetries defined in the previous section are
proportional to $\Im m (l_{mj}^{\sb*}k_{mj}^{\sb})$, see Eq.~(\ref{A2}).
Hence they measure combinations of CP phases in the MSSM.
In order to see more easily the dependence of the CP asymmetries
on the parameters, it is useful to expand:
\be
\Im m (l_{mj}^{\sb*}k_{mj}^{\sb}) =-Y_t \left[c_m~ \Im m (V_{j2}^*U_{j1}^*)
-\frac{1}{2}Y_b\sin 2\theta_{\sb}~d_m~
\Im m (V_{j2}^*U_{j2}^*e^{-i\phi_{\sb}})\right],\label{CP}
\ee
with $Y_t$ and $Y_b$ the top quark and bottom quark Yukawa couplings,
$c_1=\cos^2\theta_{\sb}$, $c_2=\sin^2\theta_{\sb}$, $d_1=1$, $d_2=-1$,
and $\theta_{\sb}$ and $\phi_{\sb}$ the mixing angle and the CP phase
of the scalar bottom system given in Appendix \ref{app:squarks}.
In general the quantity in Eq.~(\ref{CP}) can be large due to the large 
$t$- and $b$-quark Yukawa couplings. The relevant phases are
$\phi_\mu$ and $\phi_{A_b}$.
For $\phi_\mu=0$, we have $\Im m (l_{mj}^{\sb*}k_{mj}^{\sb})
\propto \sin 2\theta_{\sb} \sin\phi_{\sb}$ and from the explicit
expressions given in Appendix \ref{app:squarks}, we obtain 
$\sin 2\theta_{\sb} \sin\phi_{\sb}\propto \sin\phi_{A_b}$.
As we will see below also the asymmetries show such a $\sin\phi_{A_b}$ 
behavior and thus, their largest values are attained at 
$\phi_{A_b}=\pi/2,3\pi/2$. As $\phi_{\tilde b}$ is 
sensitive to $\phi_{A_b}$ if $|A_b|\gsim |\mu|\tan\beta$,
we need a small value for $\tan\beta$ and a large value for $|A_b|$ 
compatible with the constraint due to the tree-level
vacuum stability condition \cite{Derendinger-Savoi}.
Note that in the case where $|\mu|\tan\beta \gg |A_b|$ we have 
$\sin\phi_{\sb}\approx 0$ if $\phi_\mu=0,\pi$.

For our numerical studies we adopt the two scenarios given in Table \ref{scentab}.
In the two scenarios we have assumed the gaugino mass relation
$|M_1|=5/3\tan^2\Theta_W M_2$, with $\phi_{M_1}=0$, and we have fixed the scalar bottom masses
assuming $M_{\ti Q}>M_{\ti D}$. In scenario A the scalar bottom masses are 
heavy enough to allow for all considered decays of
$\chm_j $, Eq.~(\ref{I})--(\ref{III}), whereas the scalar bottom 
masses of scenario B are relatively light and the decay 
$\chm_j  \to  W^- \nt_1$ is not allowed. 
For the matter of simplicity, our numerical investigation is done for the first 
and second generation leptons where an influence of their Yukawa couplings
can be safely neglected.
\begin{table}[t]
\renewcommand{\arraystretch}{1.3}
\begin{center}
\begin{tabular}{|c||c||c|} \hline
Scenario & A & B \\
\hline\hline
$M_2$             & \phantom{ccc}350\phantom{ccc} &
                    \phantom{ccc}250\phantom{ccc} \\ \hline
$|\mu|$           & 310 & 140 \\ \hline
$\phi_\mu=\phi_{M_1}$        & 0   & 0   \\ \hline
$\tan\beta$       & 3   & 3   \\ \hline
$|A_b|$             & 1200 & 1000 \\ \hline
$m_{\tilde{b}_1}$ & 480 & 320 \\ \hline
$m_{\tilde{b}_2}$ & 600 & 420 \\ \hline
$m_{\tilde{\ell}_R}$ & 200 & 100 \\ \hline
$m_{\tilde{\ell}_L}$ & 220 & 120 \\ \hline
$m_{\tilde{\nu}}$ & 208.1 & 96.4 \\ \hline
$m_{\tilde{\chi}^0_1}$ & 164.3 & 80.3 \\ \hline
$m_{\tilde{\chi}^-_1}$ & 257.3 & 107.7 \\ \hline
\end{tabular}
\caption{\label{scentab}
Input parameters $M_2$, $|\mu|$, $\phi_\mu$, $\tan\beta$,
$|A_b|$, $\phi_{A_b}$, $m_{\tilde{b}_1}$, $m_{\tilde{b}_2}$,
$m_{\tilde{\ell}_R}$ and $m_{\tilde{\ell}_L}$ for scenarios
A and B.
All mass dimension parameters are given in GeV.}
\end{center}
\end{table}

\begin{figure}[t]
\setlength{\unitlength}{1mm}
\begin{center}
\begin{picture}(150,120)
\put(-53,-65){\mbox{\epsfig{figure=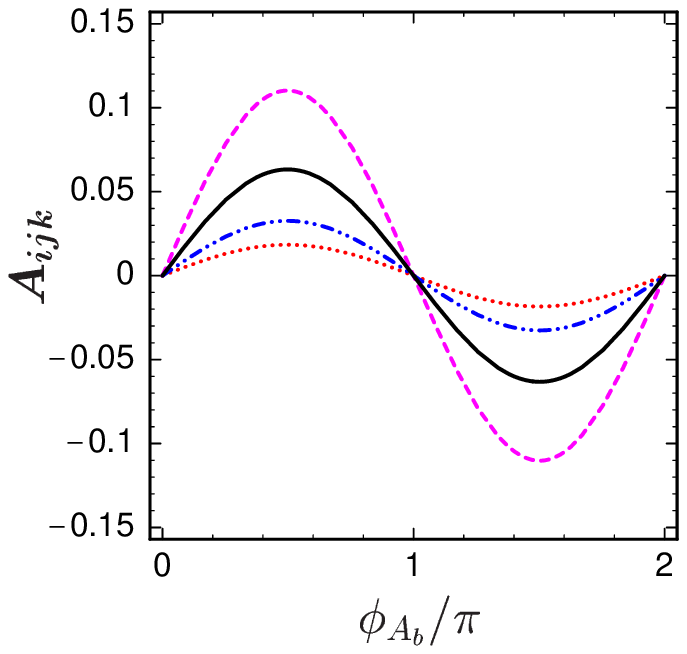,height=22.cm,width=19.4cm}}}
\put(27,-65){\mbox{\epsfig{figure=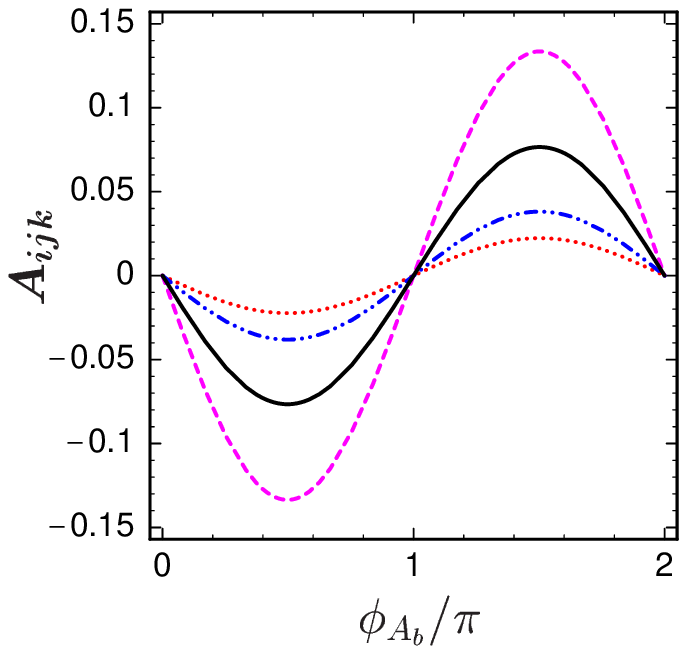,height=22.cm,width=19.4cm}}}
\put(-53,-125){\mbox{\epsfig{figure=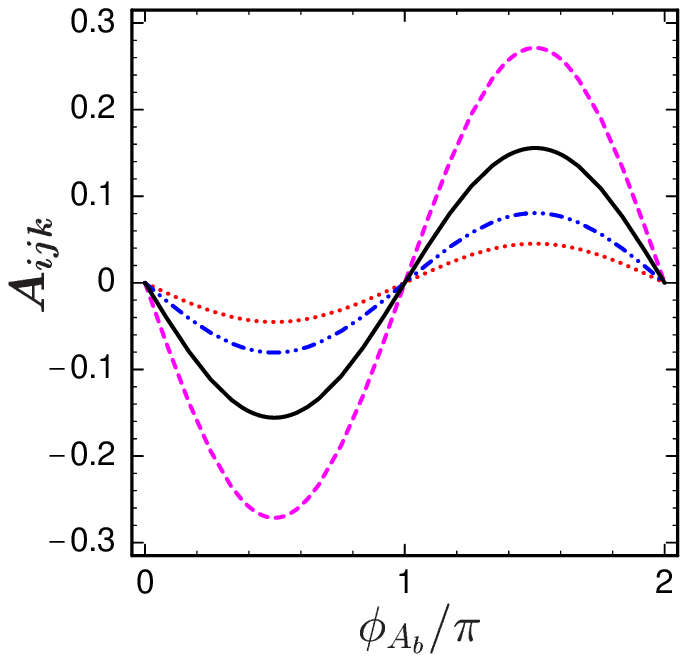,height=22.cm,width=19.4cm}}}
\put(27,-125){\mbox{\epsfig{figure=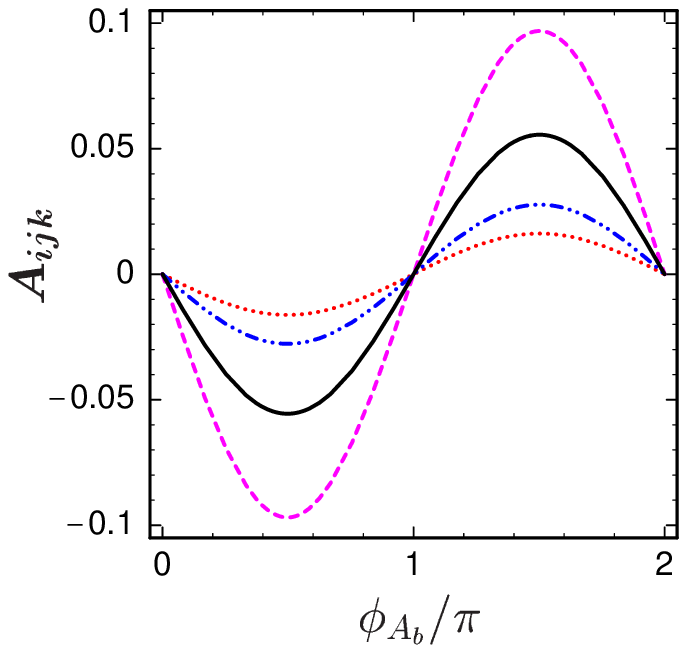,height=22.cm,width=19.4cm}}}
\put(15,129){\mbox{{\bf (a)}}}
\put(95,129){\mbox{{\bf (b)}}}
\put(15,70){\mbox{{\bf (c)}}}
\put(95,69){\mbox{{\bf (d)}}}
\end{picture}
\end{center}
\vspace{-2cm}
\caption{CP asymmetries
$A_{ijk}$ which are based on the triple products (a)
$({\bf p}_{\ell^-_i}{\bf p}_b{\bf p}_t)$,
(b) $({\bf p}_{\ell^-_i}{\bf p}_{l^+}{\bf p}_b)$,
(c) $({\bf p}_{\ell^-_i}{\bf p}_{c}{\bf p}_t)$
and (d) $({\bf p}_{\ell^-_i}{\bf p}_{c}{\bf p}_{s})$
for the decays
$\tilde{b}_1\to t\tilde{\chi}^-_1$,
$t\to bl^+\nu \, (bc\bar{s})$ and
$\tilde{\chi}^-_1 \to \ell^-_i\bar\nu \tilde{\chi}^0_1$,
as a function of $\phi_{A_b}$.
The lepton $\ell^-_1$ ($\ell^-_2,\ell^-_3$) stems from the decay
$\tilde{\chi}^-_1\to\ell^-_1\bar{\tilde{\nu}} \to \ell^-_1\bar\nu\tilde{\chi}^0_1$
($\tilde{\chi}^-_1\to\tilde{\ell}^-_R\bar\nu \to \ell^-_2\bar\nu\tilde{\chi}^0_1$,
$\tilde{\chi}^-_1\to W^-\tilde{\chi}^0_1  \to \ell^-_3\bar\nu\tilde{\chi}^0_1$).
The corresponding asymmetries are shown as dashed lines 
(dotted lines, dashdotdotted lines).
The solid lines represent the combined asymmetries in Eq.~(\ref{sumasy}).
The MSSM parameters are for scenario A of Table \ref{scentab}.}
\label{fig:fig1}
\end{figure}

In Fig. \ref{fig:fig1} we show the CP asymmetries that are based
on the triple products, (\ref{Asymbi})--(\ref{Asymci}), in the decays
$\tilde{b}_1\to t\tilde{\chi}^-_1$,
$t\to bl^+\nu \, (bc\bar{s})$ and
$\tilde{\chi}^-_1 \to \ell^-_i \bar{\nu} \tilde{\chi}^0_1$
as a function of $\phi_{A_b}$.
Fig. \ref{fig:fig1}(a) shows the three CP asymmetries $A_{\ell_i^-bt}$ that are based
on the triple products in Eq.~(\ref{Asymbi}) associated
with the three different decay chains
$\tilde{\chi}^-_1\to\ell^-_1 \bar{\tilde{\nu}} \to
\ell^-_1\bar{\nu} \tilde{\chi}^0_1$ (dashed line),
$\tilde{\chi}^-_1\to \tilde{\ell}^-_R\nu \to
\ell^-_2\bar{\nu} \tilde{\chi}^0_1$ (dotted line)
and $\tilde{\chi}^-_1\to W^-\tilde{\chi}^0_1 \to
\ell^-_3\bar{\nu} \tilde{\chi}^0_1$
(dashdotdotted line).
Fig. \ref{fig:fig1}(a) also shows the CP asymmetry
$A_{\ell^-bt}$ (solid line), Eq.~(\ref{sumasy}),
where it is not necessary to distinguish the leptons from the different decay
chains of the chargino.
The asymmetry $A_{\ell^-_1 b t}$ is the largest one with
a maximum value of about $11\%$.
The CP asymmetries $A_{\ell^-_2 b t}$ and
$A_{\ell^-_3 b t}$ have an additional phase space
factor and are therefore suppressed compared to $A_{\ell^-_1 b t}$.

We now estimate the number of scalar bottoms $\tilde{b}_1$ necessary to observe the
CP asymmetries for a given number of standard deviations $\mathcal{N}_\sigma$ by 
\be
N_{\tilde{b}_1}=\frac{{\mathcal{N}_\sigma}^2}
{A_{ijk}^2 \, BR(\tilde{b}_1\to t\tilde{\chi}^-_1)
(\sum_{\ell =e,\mu} BR(\tilde{\chi}^-_1 \to \ell^-_i \bar{\nu} \tilde{\chi}^0_1))
(\sum_f BR(W^+ \to f))},
\label{estimate}
\ee
where $f$ denotes the final state of the $W^+$ decay considered, i.e.
$f= u\bar{d}, c\bar{s}$, or $l^+\nu_l$, $l=e,\mu$.
We calculate the branching ratios of $\ti b_1$ using
the formulae of the second paper in \cite{Bartl:2003he}.
For scenario A we obtain $BR(\tilde{b}_1\to t\tilde{\chi}^-_1)=4.9\%$.
Purely for the sake of simplicity, we calculate the chargino branching ratios
$BR(\tilde{\chi}^-_1 \to \ell^-_i \bar{\nu} \tilde{\chi}^0_1)$
assuming that scalar tau mixing can be neglected
and that the lighter scalar leptons have a common mass $m_{\ti\ell_R}$,
the heavier scalar leptons have a common mass $m_{\ti\ell_L}$
and the scalar neutrinos have a common mass given by
\be
m_{\tilde{\nu}_\ell}=\sqrt{m_{\tilde{\ell}_L}^2+m_Z^2\cos 2\beta \cos^2\theta_W}~.
\ee
This means that the partial decay widths $\Gamma(\ti\chi^-_1\to \ell^- \bar{\ti\nu}_\ell)$
are equal for $\ell=e,\mu,\tau$. The same holds for the partial decay widths
$\Gamma(\ti\chi^-_1\to \ti\ell^-_R\bar{\nu}_\ell)$ and
$\Gamma(\ti\ell^-_R\to \ti\chi^0_1 \ell^-)$.
Then we obtain
$\sum_{\ell=e,\mu}BR(\tilde{\chi}^-_1 \to \ell^-_i \bar{\nu} \tilde{\chi}^0_1)=
(31.3\%, 30.7\%, 1.5\%)$ corresponding to the three different decay
chains of $\ti\chi^-_1$, (\ref{I})--(\ref{III}).
The values of the branching ratios of the $W$ boson are given
by $BR(W^+ \to \sum_l l^+ \nu)=21.4\%$ ($l=e,\mu$),
$BR(W^+ \to \sum_q q\bar{q}'=68\%)$ and $BR(W^+ \to c\bar{s}=32\%)$ \cite{particledata}.
For an observation of the CP asymmetry $A_{\ell^-_1 b t}$ at the 3-$\sigma$ level,
at least $7.1\cdot10^4$ scalar bottoms have to be produced.
The required number of scalar bottoms in order to measure the asymmetry
$A_{\ell^- b t}=6.4\%$ ($\phi_{A_b}=0.5\pi$) at the 3-$\sigma$ level
is $1\cdot 10^5$.

In Fig. \ref{fig:fig1}(b) we plot the CP
asymmetries that are based on the triple products
$({\bf p}_{\ell^-_i}{\bf p}_{l^+}{\bf p}_b)$, Eq.~(\ref{Asymli}), as
a function of $\phi_{A_b}$. For the same reason as above the largest asymmetry is due to
the chargino decay chain $\tilde{\chi}^-_1\to\ell^-_1 \bar{\tilde{\nu}} \to
\ell^-_1\bar{\nu} \tilde{\chi}^0_1$, (\ref{I}).
Its maximal value of about $13\%$ is reached at
$\phi_{A_b}=0.5\pi$ and the number of scalar bottoms necessary to
measure $A_{\ell^-_1 l^+ b}$ at the 3-$\sigma$ level is about $1.5\cdot 10^5$.
Fig. \ref{fig:fig1}(c) shows the CP asymmetries that are
based on $({\bf p}_{\ell^-_i}{\bf p}_{c}{\bf p}_t)$ as a function
of $\phi_{A_b}$. The asymmetries shown in Fig. \ref{fig:fig1}(c)
are more than twice as large as the asymmetries shown in Fig.~\ref{fig:fig1}.
Their relative magnitudes can be attributed ($i$) to the different sensitivity
factors of the top quark polarization which is $\alpha_l=1$ for
the asymmetries in Fig.\ref{fig:fig1}(b),(c),(d) and $\alpha_b\simeq 0.4$
for the asymmetries in Fig. \ref{fig:fig1}(a), and ($ii$) to the different 
3-vectors involved in the triple products:
for Figs.\ref{fig:fig1}(a) and \ref{fig:fig1}(c) it is ${\bf p}_t$, while 
for Figs. \ref{fig:fig1}(b) and \ref{fig:fig1}(d) it is the 3-vector of any 
of the decay products of the $t$-quark, which is always smaller or at most equal
in magnitude than $|{\bf p}_t|$.
For $\phi_{A_b}=0.5\pi$ the CP asymmetry $A_{\ell^-_1 c t}$ is about $27\%$,
which means that $2.5\cdot 10^4$ scalar bottoms are necessary
for its measurement at 3-$\sigma$.
The combined asymmetry in Eq.~(\ref{sumasy}) can be as large as about $16\%$
and the appropriate number of scalar bottoms to probe it at the 3-$\sigma$ level is
$3.6\cdot 10^4$.
In Fig. \ref{fig:fig1}(d) we plot the CP asymmetries which are based on the triple 
products $({\bf p}_{\ell^-_i}{\bf p}_{c}{\bf p}_{\bar{s}})$. For $\phi_{A_b}=0.5\pi$ 
the asymmetry is $A_{\ell^-_1 c \bar{s}}$ of about $10\%$ and at least $1.9\cdot 10^5$ 
scalar bottoms are required for its measurement.

\begin{figure}[t]
\setlength{\unitlength}{1mm}
\begin{center}
\begin{picture}(180,180)
\put(-52,-90){\mbox{\epsfig{figure=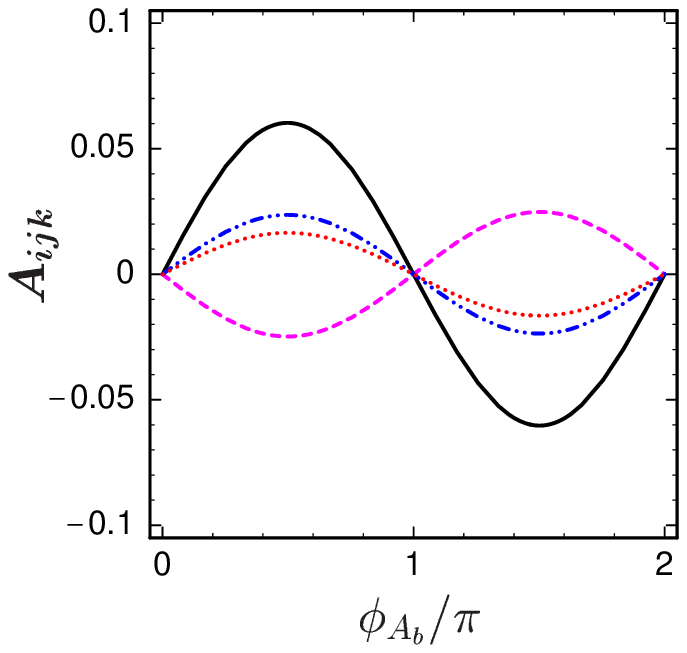,height=31.cm,width=27.4cm}}}
\end{picture}
\end{center}
\vspace{-11.3cm}
\caption{CP asymmetries
$A_{ijk}$ that are based on the triple products
$({\bf p}_{W^-}{\bf p}_c{\bf p}_t)$ (solid line),
$({\bf p}_{W^-}{\bf p}_{c}{\bf p}_{\bar{s}})$ (dotted line),
$({\bf p}_{W^-}{\bf p}_b{\bf p}_t)$ (dashed line)
and $({\bf p}_{W^-}{\bf p}_{l^+}{\bf p}_b)$ (dashdotdotted line)
for the process
$\tilde{b}_1\to t\tilde{\chi}^-_1$,
$t\to bl^+\nu \, (bc\bar{s})$ and
$\tilde{\chi}^-_1\to W^-\tilde{\chi}^0_1 \to \bar{c}s \tilde{\chi}^0_1$
as a function of $\phi_{A_b}$.
The MSSM parameters are given in Table \ref{scentab} (scenario A).}
\label{fig:fig2}
\end{figure}

In Fig. \ref{fig:fig2} we show the CP asymmetries
that are based on the triple products $({\bf p}_{W^-}{\bf p}_c{\bf p}_t)$,
$({\bf p}_{W^-}{\bf p}_{c}{\bf p}_{s})$,
$({\bf p}_{W^-}{\bf p}_b{\bf p}_t)$
and $({\bf p}_{W^-}{\bf p}_{l^+}{\bf p}_b)$ as
a function of $\phi_{A_b}$ for scenario A given in Table \ref{scentab}.
The momentum vector ${\bf p}_{W^-}$ involved in the triple
products is that of the $W$ boson stemming from the
decay $\tilde{\chi}^-_1\to W^-\tilde{\chi}^0_1$.
The largest asymmetry $A_{W^- c t}$ attains its maximum
value of about $6\%$ at $\phi_{A_b}=0.5\pi$.
For the theoretical estimate of the number of scalar bottoms
necessary to observe this asymmetry we replace
$\sum_{\ell=e,\mu} BR(\tilde{\chi}^-_1 \to \ell^-_i\bar{\nu} \tilde{\chi}^0_1)$
by $BR(\ti\chi^-_1\to W^-\ti\chi_1^0)\cdot
\sum_q BR(W^-\to \bar{q} q')= 4.8\%$ in Eq.~(\ref{estimate}).
We then obtain that $1.1 \cdot 10^6$ scalar bottoms
are required for a 3-$\sigma$ evidence.

\begin{figure}[t]
\setlength{\unitlength}{1mm}
\begin{center}
\begin{picture}(150,120)
\put(-53,-65){\mbox{\epsfig{figure=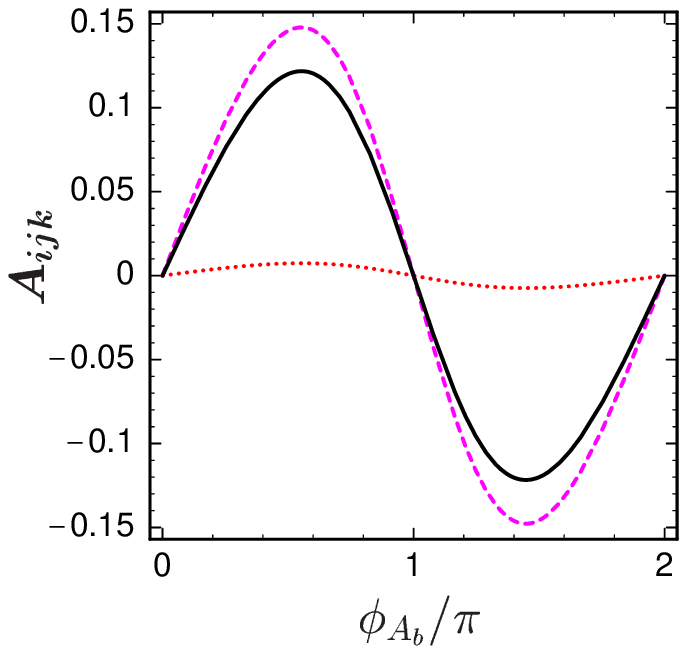,height=22.cm,width=19.4cm}}}
\put(27,-65){\mbox{\epsfig{figure=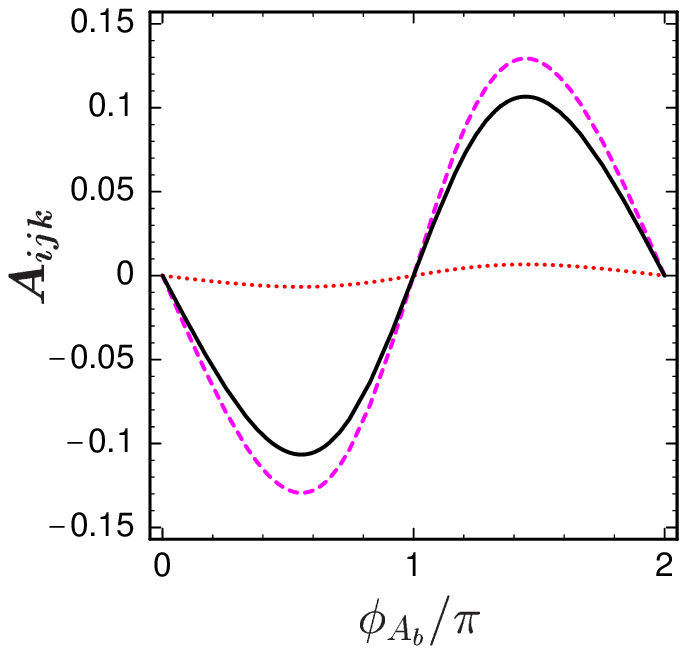,height=22.cm,width=19.4cm}}}
\put(-53,-125){\mbox{\epsfig{figure=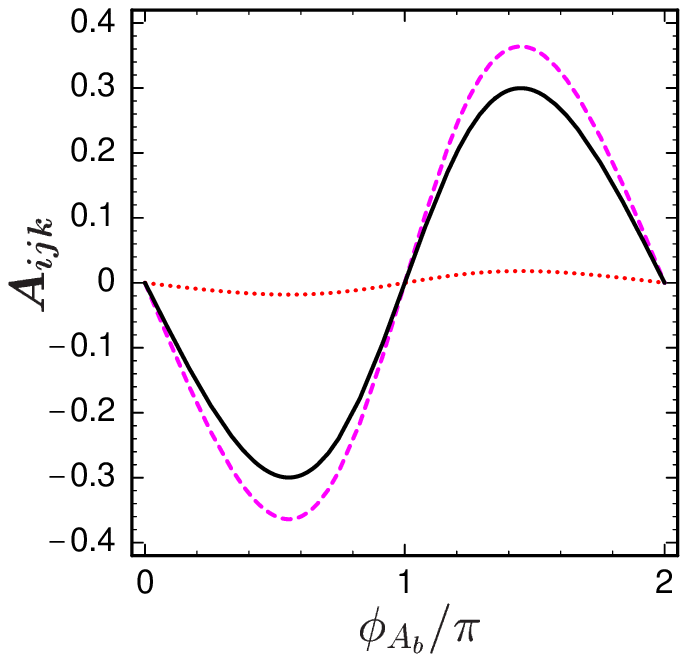,height=22.cm,width=19.4cm}}}
\put(27,-125){\mbox{\epsfig{figure=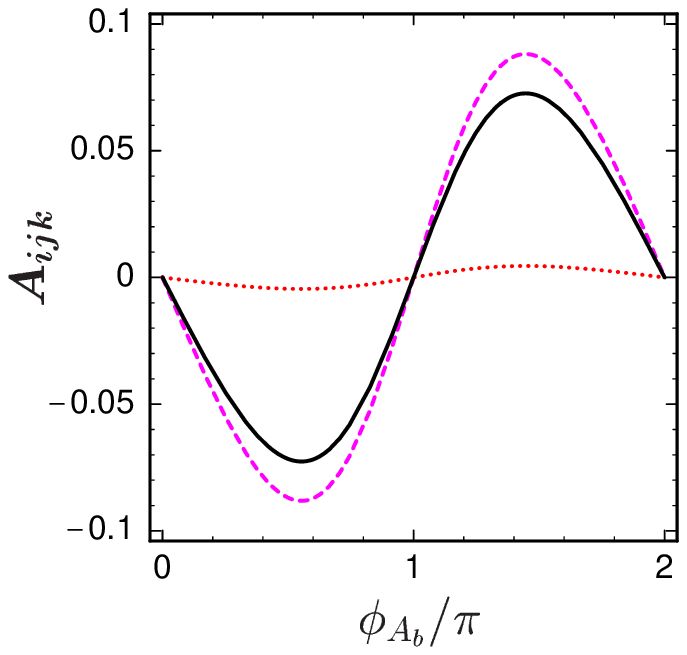,height=22.cm,width=19.4cm}}}
\put(15,129){\mbox{{\bf (a)}}}
\put(95,129){\mbox{{\bf (b)}}}
\put(15,70){\mbox{{\bf (c)}}}
\put(95,69){\mbox{{\bf (d)}}}
\end{picture}
\end{center}
\vspace{-2cm}
\caption{CP asymmetries
$A_{ijk}$ that are based on the triple products (a)
$({\bf p}_{\ell^-_i}{\bf p}_b{\bf p}_t)$,
(b) $({\bf p}_{\ell^-_i}{\bf p}_{l^+}{\bf p}_b)$,
(c) $({\bf p}_{\ell^-_i}{\bf p}_{c}{\bf p}_t)$
and (d) $({\bf p}_{\ell^-_i}{\bf p}_{c}{\bf p}_{s})$
for the decays
$\tilde{b}_1\to t\tilde{\chi}^-_1$,
$t\to bl^+\nu \, (bc\bar{s})$ and
$\tilde{\chi}^-_1 \to \ell^-_i\bar\nu \tilde{\chi}^0_1$,
as a function of $\phi_{A_b}$.
The lepton $\ell^-_1$ ($\ell^-_2$) stems from the decay
$\tilde{\chi}^-_1\to\ell^-_1 \bar{\tilde{\nu}}\to \ell^-_1\bar\nu\tilde{\chi}^0_1$
($\tilde{\chi}^-_1\to\tilde{\ell}^-_R\bar\nu \to \ell^-_2\bar\nu\tilde{\chi}^0_1$).
The corresponding asymmetries are shown as dashed lines (dotted lines).
The solid lines represent the combined asymmetries in Eq.~(\ref{sumasy}).
The MSSM parameters are for scenario B of Table \ref{scentab}.}
\label{fig:fig3}
\end{figure}

In Fig. \ref{fig:fig3} the CP asymmetries for scenario B
of Table \ref{scentab} are displayed.
In this case the decay $\tilde{\chi}^-_1\to W^-\tilde{\chi}^0_1$ is
kinematically not accessible. We plot the CP asymmetries for the decay chains
$\tilde{\chi}^-_1\to\ell^-_1\bar{\tilde{\nu}} \to \ell^-_1\bar{\nu} \tilde{\chi}^0_1$
and $\tilde{\chi}^-_1\to\tilde{\ell}^-_R \bar{\nu} \to \ell^-_2\bar{\nu} \tilde{\chi}^0_1$
as a function of $\phi_{A_b}$.
For the branching ratios we obtain
$BR(\tilde{b}_1\to t\tilde{\chi}^-_1)=7.2\%$,
$\sum_{\ell=e,\mu} BR(\tilde{\chi}^-_1 \to \ell^-_1\bar{\nu} \tilde{\chi}^0_1)=
54.3\%$ and
$\sum_{\ell=e,\mu} BR(\tilde{\chi}^-_1 \to \ell^-_2\bar{\nu} \tilde{\chi}^0_1)=12.4\%$
in scenario B.
Fig. \ref{fig:fig3}(a) shows the CP asymmetries which are based
on the triple products given in Eq.~(\ref{Asymbi}).
The largest asymmetry results from the triple product
$({\bf p}_{\ell^-_1} {\bf p}_b {\bf
p}_{t})$ where the lepton $\ell^-_1$ originates from
the first step of the decay chain
$\tilde{\chi}^-_1\to\ell^-_1\bar{\tilde{\nu}} \to \ell^-_1\bar{\nu} \tilde{\chi}^0_1$,
and its maximum value is of about $15\%$. For its
measurement (at 3-$\sigma$) $16\cdot 10^4$ scalar bottoms are required.
The CP asymmetry that is based on $({\bf p}_{\ell^-_2} {\bf p}_b {\bf
p}_{t})$, where the lepton $\ell^-_2$ comes from the decay chain
$\tilde{\chi}^-_1\to\tilde{\ell}^-_R \bar{\nu} \to \ell^-_2\bar{\nu} \tilde{\chi}^0_1$
is phase space suppressed. Due to the large branching ratio of
$\tilde{\chi}^-_1 \to \ell^-_1\bar{\nu} \tilde{\chi}^0_1$
the combined asymmetry, Eq.~(\ref{sumasy}),
is about $12\%$, therefore $1.9\cdot 10^4$
scalar bottoms would be necessary for a measurement at the
3-$\sigma$ level. In Fig. \ref{fig:fig3}(b) we plot the
CP asymmetries that are based on the triple products defined
in Eq.~(\ref{Asymli}). The largest asymmetry $A_{\ell^-_1 l^+ b}$
reaches its maximum value of about $13\%$ at $\phi_{A_b}=1.5\pi$.
Fig. \ref{fig:fig3}(c) shows the CP asymmetry formed with
the triple products $({\bf p}_{\ell^-_i}{\bf p}_{c}{\bf p}_t)$.
As expected, the asymmetry $A_{\ell^-_1 c t}$ is the largest and its maximum value 
is of about $36\%$. In this case $5.5\cdot 10^3$ scalar bottoms
are necessary for a measurement of $A_{\ell^-_1 c t}$ at the 3-$\sigma$ level.
The CP asymmetry, where it is not necessary to distinguish
from which $\tilde{\chi}^-_1$ decay chain the lepton originates,
reaches a maximum of about $30\%$. In this case the production
of $6.7 \cdot 10^3$ scalar bottoms is necessary to probe
the asymmetry $A_{\ell^- c t}$ at 3-$\sigma$.
In Fig. \ref{fig:fig3}(d) the CP asymmetries that are based on the triple products
$({\bf p}_{\ell^-_i}{\bf p}_{c}{\bf p}_{s})$ are displayed.
The maximum of $A_{\ell^-_1 c s}$ is about $9\%$, which means
that $1.1 \cdot 10^5$ scalar bottoms are necessary to determine (at 3-$\sigma$)
that the asymmetry is non-zero.

\begin{figure}[t]
\vspace{1cm}
\setlength{\unitlength}{1mm}
\begin{center}
\begin{picture}(180,167)
\put(-39,-90){\mbox{\epsfig{figure=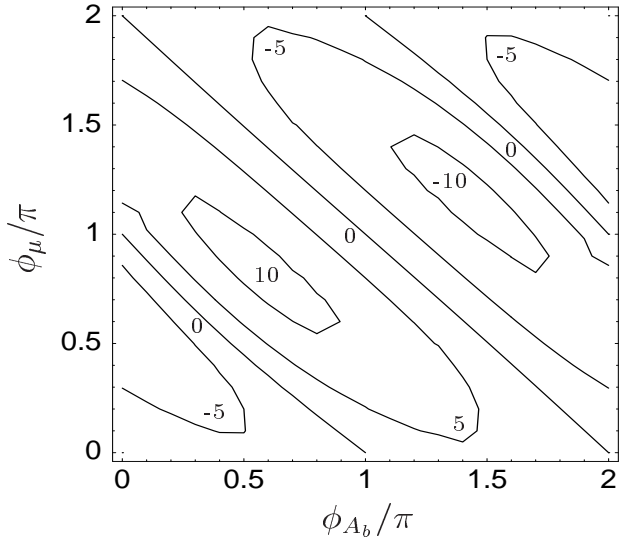,height=31cm,width=24.4cm}}}
\end{picture}
\end{center}
\vspace{-11.5cm}
\caption{Contours in the $\phi_{\mu}$-$\phi_{A_b}$ plane of the combined 
asymmetry, Eq.~(\ref{sumasy}), which is based on 
$({\bf p}_{\ell^-}{\bf p}_{c}{\bf p}_t)$. We take $m_{\tilde{\ell}_L}=140$~GeV,
$m_{\tilde{\ell}_R}=110$~GeV and $\tan\beta=10$, the other parameters are as 
in scenario B of Table \ref{scentab}.}
\label{fig:fig4}
\end{figure}

Fig. \ref{fig:fig4} shows the contours of the combined asymmetry
$A_{\ell^- c t}$, Eq.~(\ref{sumasy}), in the
$\phi_{\mu}$-$\phi_{A_b}$ plane.
The slepton masses are $m_{\tilde{\ell}_L}=140$~GeV and $m_{\tilde{\ell}_R}=110$~GeV,
$\tan\beta=10$ and the other parameters are as given as in scenario B
of Table \ref{scentab}. For the
scenario chosen, the first term of Eq.~(\ref{CP}) is small compared to the second term
because $\cos\theta_{\ti b} \ll \sin\theta_{\ti b}$.
Hence, the behavior of the asymmetry is given by the second term of Eq.~(\ref{CP}),
which is small in the $\phi_{\mu}$-$\phi_{A_b}$ plane where
$\phi_\mu + \phi_{A_b}\approx 0, \pi$ because
there $\phi_{\ti b}-\arg[U^*_{12}V^*_{12}]\approx 0, \pi$ resulting
in a cancellation of the two terms in Eq.~(\ref{CP}).
For CP phases of $\phi_{\mu}\approx 0.8\pi$ and
$\phi_{A_b}\approx 0.6\pi$ the
asymmetry reaches its maximum of about $11\%$.

\section{Summary \label{concl}}

We have proposed various T-odd asymmetries in the decay $\sb_m \to t \ti\chi_j^-$,
which are based on triple product correlations that involve the polarization vectors
of $t$ and $\ti\chi_j^-$. The distributions of their decay products depend on the 
polarizations of $t$ and $\ti\chi_j^-$.
For the $\ti\chi_j^-$ decay into a leptonic final state $\ell^- \bar{\nu} \ti\chi^0_1$
we have considered the three possible decay chains
$\ti\chi_j^- \to \ell^-\bar{\sn} \to \ell^- \bar{\nu} \ti\chi^0_1$,
$\ti\chi_j^- \to {\tilde\ell}^-_n \to\ell^- \bar{\nu} \ti\chi^0_1$ and
$\ti\chi_j^- \to W^-\nt_1 \to \ell^- \bar{\nu} \ti\chi^0_1$.
We have also considered the 2-body decay $\ti\chi_j^- \to W^-\chi^0_1$, where the
$W$ boson decays hadronically.
The proposed T-odd asymmetries are proportional to the product of
left- and right-couplings $t~\ti b_m \ti\chi_k^-$
and are non-vanishing due to non-zero phases $\phi_\mu$ and/or $\phi_{A_b}$.
Since scalar bottom mixing can be large these asymmetries
will allow us to determine the CP violating phase $\phi_{A_b}$, which is not easily
accessible otherwise. We have also pointed out that true CP violating asymmetries
can be obtained by summing the T-odd asymmetries that arise in the decays
$\sb_m\to\ti\chi^-_j t$ and ${\bar{\tilde b}}_m\to\ti\chi^+_j \bar{t}$.
In this case an identification of the charges of the involved particles is not
necessary.

In a numerical study we have presented results of these asymmetries
for the decay $\sb_1 \to t \ti\chi_1^-$.
The asymmetry $A_{\ell_1^-ct}$, which is based on the triple
product $({\bf p}_{\ell^-_1} {\bf p}_{c} {\bf p}_{t})$, is the largest one
and its magnitude can be of the order $40\%$.
We have also defined the asymmetry $A_{\ell^- ct}$, Eq.~(\ref{sumasy}), which is
based on $({\bf p}_{\ell^-} {\bf p}_{c} {\bf p}_{t})$, and where
it is not necessary to distinguish between the different leptonic $\tilde{\chi}^-_1$ 
decay chains. We have found that this asymmetry can go up to $30\%$.
By making a theoretical estimate of the number
of $\ti b_1$ necessary to observe the T-odd asymmetries we
have found that a $\ti b_1$ production rate of $O(10^3)$
will be necessary to observe some of the proposed asymmetries,
which should be possible at the LHC or at a future linear collider.

\section*{Acknowledgements}

This work is supported by the 'Fonds zur F\"orderung der
wissenschaftlichen Forschung' (FWF) of Austria, project. No. P18959-N16.

\section*{Appendix}
\begin{appendix}

\section{Scalar bottom masses and mixing}
\label{app:squarks}

The left-right mixing of the scalar bottoms is described by a
hermitian $2 \times 2$ mass matrix which in the basis
$(\tilde{b}_L,\tilde{b}_R)$ reads
\bee{eq:stopmass}
{\mathcal{L}}_M^{\sb}= -(\sb_L^{\dagger},\, \sb_R^{\dagger})
\left(\begin{array}{ccc}
M_{\sb_{LL}}^2 & e^{-i\phi_{\sb}}|M_{\sb_{LR}}^2|\\[5mm]
e^{i\phi_{\sb}}|M_{\sb_{LR}}^2| & M_{\sb_{RR}}^2
\end{array}\right)\left(
\begin{array}{ccc}
\sb_L\\[5mm]
\sb_R \end{array}\right),
\eee
where
\begin{eqnarray}
M_{\sb_{LL}}^2 & = & M_{\tilde Q}^2+(-\frac{1}{2}+\frac{1}{3}\sin^2\Theta_W)
\cos2\beta \ m_Z^2+m_b^2 ,\label{eq:mll} \\[3mm]
M_{\sb_{RR}}^2 & = & M_{\tilde D}^2-\frac{1}{3}\sin^2\Theta_W\cos2\beta \
m_Z^2+m_b^2 ,\label{eq:mrr}\\[3mm]
M_{\sb_{RL}}^2 & = & (M_{\sb_{LR}}^2)^{\ast}=
m_b(A_b-\mu^{\ast}
\tan\beta), \label{eq:mlr}
\end{eqnarray}
\begin{equation}
\phi_{\sb}  = \arg\lbrack A_b-\mu^{\ast}\tan\beta\rbrack ,
\label{eq:phtau}
\end{equation}
where $\tan\beta=v_2/v_1$ with $v_1 (v_2)$ being the vacuum
expectation value of the Higgs field $H_1^0 (H_2^0)$,
$m_b$ is the mass of the bottom quark and
$\Theta_W$ is the weak mixing angle, $\mu$ is the Higgs--higgsino mass parameter
and $M_{\ti Q}$,
$M_{\ti D}, A_t$ are the soft SUSY--breaking parameters of the scalar bottom system.
The mass eigenstates $\sb_i$ are $(\ti b_1, \ti b_2)=
(\sb_L, \sb_R) {\mathcal{R}^{\sb}}^T$ with
\begin{equation}
\mathcal{R}^{\sb}=\left( \begin{array}{ccc}
e^{i\phi_{\sb}}\cos\theta_{\sb} &
\sin\theta_{\sb}\\[5mm]
-\sin\theta_{\sb} &
e^{-i\phi_{\sb}}\cos\theta_{\sb}
\end{array}\right),
\label{eq:rtau}
\end{equation}
with
\begin{equation}
\cos\theta_{\sb}=\frac{-|M_{\sb_{LR}}^2|}{\sqrt{|M_{\sb _{LR}}^2|^2+
(m_{\sb_1}^2-M_{\sb_{LL}}^2)^2}},~
\sin\theta_{\sb}=\frac{M_{\sb_{LL}}^2-m_{\sb_1}^2}
{\sqrt{|M_{\sb_{LR}}^2|^2+(m_{\sb_1}^2-M_{\sb_{LL}}^2)^2}}.
\label{eq:thtau} 
\end{equation}
The mass eigenvalues are
\begin{equation}
 m_{\sb_{1,2}}^2 = \frac{1}{2}\left((M_{\sb_{LL}}^2+M_{\sb_{RR}}^2)\mp
\sqrt{(M_{\sb_{LL}}^2 - M_{\sb_{RR}}^2)^2 +4|M_{\sb_{LR}}^2|^2}\right).
\label{eq:m12}
\end{equation}
%

\section{Lagrangian and couplings \label{app:lagrange}}

\noi
The parts of the Lagrangian, necessary to calculate
the  decay rates of
$\ti b_m \to \tilde{\chi}^-_jt$ with the subsequent decays
$\tilde{\chi}^-_j\to \ell^-\bar{\nu} \tilde\chi^0_1$ are
\baq{eq:lagrangian}
{\cal{L}}_{t\tilde{b}\chi^+}&=&
g\,\bar{t}\,(l_{mj}^{\ti b}\,\PR + k_{mj}^{\ti b}\,\PL) \ti\chi^+_j \ti b_m
+{\rm h.c.}~,\\
{\cal L}_{\ell \tilde{\nu}\tilde{\chi}^+}&=&
 g\,\bar{\ell}\,(k_{j}^{\ti \nu}\,P_L + l_{j}^{\ti \nu}\,P_R) \ti\chi^{+C}_j \ti\nu_\ell
+ {\rm h.c.}~,\\
{\cal L}_{\nu\tilde{\ell}\tilde{\chi}^+}&=&
g~ l_{nj}^{\ti\ell}~ \bar{\nu_\ell}~ P_R~\ti\chi^+_j~ \ti \ell_n
+ {\rm h.c.}~,\\
{\cal L}_{W^-\tilde{\chi}^+\tilde{\chi}^0}&=&
g W^-_{\mu} \overline{\tilde{\chi}^0_k}\gamma^{\mu}(O^L_{kj} P_L+O^R_{kj} P_R)
\tilde{\chi}^+_j+{\rm h.c.}~,\\
{\cal L}_{\ell\tilde{\ell}\tilde{\chi}^0}&=&
g\,\bar{\ell}\,(a_{nk}^{\ti\ell}\,\PR + b_{nk}^{\ti\ell}\,\PL)\,\nt_k\,\ti\ell_n
+ {\rm h.c.}~,\\
{\cal L}_{\nu\tilde{\nu}\tilde{\chi}^0} &=&
g~ f_{Lk}^\nu \bar{\nu_\ell} P_R \nt_k~\ti\nu_\ell
+ {\rm h.c.}~,
\eaq
where the couplings are defined as
\bee{eq:couplsbott}
l_{mj}^{\ti b} = -{\mathcal R}^{\ti b *}_{m1} U_{j1}+
Y_b {\mathcal R}^{\ti b *}_{m2} U_{j2}~, \qquad
k_{mj}^{\ti b} ={\mathcal R}^{\ti b *}_{m1}\,Y_t\,V_{j2}^{\ast}~,
\eee
\bee{eq:snucoupl}
l_{j}^{\ti\nu} = -V_{j1}~, \qquad
k_{j}^{\ti\nu} = Y_\ell U^*_{j2}~,
\eee
\bee{eq:albl}
a_{nk}^{\ti\ell}={\mathcal R}^{\ti\ell*}_{n1} f_{Lk}^\ell
+{\mathcal R}^{\ti\ell*}_{n2} h_{Rk}^\ell~,\qquad
b_{nk}^{\ti\ell}={\mathcal R}^{\ti\ell*}_{n1} h_{Lk}^\ell
+{\mathcal R}^{\ti\ell*}_{n2} f_{Rk}^\ell~,
\eee
\be
f_{Lk}^\ell & = &
{1\over \rzw}\Bigl( N_{k2}+\tan\tW N_{k1}\Bigr)\ ,
\hspace{-1cm} \label{eq:fLkl}\nonumber \\
f_{Rk}^\ell & = &
- {\rzw}\tan\tW N^{\ast}_{k1}\ ,\nonumber \\
h_{Rk}^\ell & = &(h_{Lk}^l)^*=-Y_\ell N_{k3}~,
\nonumber \\
f_{Lk}^\nu &=& {1\over \rzw}\Bigl(\tan\tW N_{k1}- N_{k2}\Bigr)~,
\ee
\bee{eq:ll}
l_{nj}^{\ti\ell} = -{\mathcal R}^{\ti\ell*}_{n1} U_{j1}+
Y_\ell {\mathcal R}^{\ti\ell*}_{n2} U_{j2}~,
\eee
\bee{eq:OLOR}
O^L_{kj}=-\frac{1}{\sqrt{2}} N_{k4}V^*_{j2} + N_{k2}V^*_{j1}~,\qquad
O^R_{kj}=\frac{1}{\sqrt{2}} N^*_{k3}U_{j2} + N^*_{k2}U_{j1}~,
\eee
where in the above equations $U$ and $V$ are the unitary $2\times2$ mixing matrices
that diagonalize the chargino mass matrix ${\mathcal M}_C$,
$U^{\ast}{\mathcal M}_C V^{-1}=
{\rm diag}(m_{\chi_1},m_{\chi_2})$,
$N_{ij}$ is the complex unitary $4\times 4$ matrix which diagonalizes
the neutral gaugino-higgsino mass matrix $Y_{\alpha\beta}$,
$N_{i \alpha}^*Y_{\alpha\beta}N_{k\beta}^{\ast}=
m_{\chi^0_i}\delta_{ik}$,
in the basis ($\tilde{B},
\tilde{W}^3, $ $ \tilde{H}^0_1, \tilde{H}^0_2$) \cite{guha},
${\mathcal R}^{\ti\ell}$ is the mixing matrix in the slepton sector
(see for instance \cite{Bartl:2002uy}) and the Yukawa couplings are given
by $Y_t=m_t/(\sqrt{2} m_W \sin\beta), Y_b=m_b/(\sqrt{2} m_W \cos\beta)$
and $Y_\ell=m_\ell/(\sqrt{2} m_W \cos\beta)$, with $m_W$ being the mass of
the $W$ boson.

\section{Phase space and kinematics\label{app:phasespace}}

We will work in the rest frame of $\tilde b_m$ and we fix the coordinate
system so that the chargino momentum ${\bf p}_{\chi_j}$
points along the $Z$-axis.

\noi
\underline{\it Phase space element of the decay $\tilde b_m\to\ti\chi_j^-t$:}
\bee{eq:phisbott}
d\Phi_{\tilde b_m }=
\frac{|{\bf p}_t|}{4\pi m_{\ti b_m}}~,\qquad
|{\bf p}_t|=
\frac{\lambda^{\frac{1}{2}}(m^2_{\ti b_m},m^2_t,m^2_{\chi_j})}{2 m_{\ti b_m}}~,
\eee
where $\lambda(x,y,z)=x^2+y^2+z^2-2(x y+x z+y z)$.

\noi
\underline{\it Phase space elements of the top decays (\ref{t}):}

\noi
The phase space element of the top decay $t\to b W^+$ is given as
\be
d\Phi_t^b =
\frac{E_b^2}{2(m_t^2-m_W^2)}~\frac{d\Omega_b}{(2\pi )^2}~,
\qquad E_b= \frac{m_t^2-m_W^2}{2(E_t+|{\bf p}_t|~c_b)}~.
\label{eq:pstopW}
\ee
The phase space element of the top decay $t\to b l^+\nu$ reads
\be
d\Phi_t^l =\frac{1}{2\pi }\,d\Phi_t^b\, d\Phi_W~,
\label{eq:pstopl}
\ee
where we used the narrow width approximation for the $W$ boson
propagator. $d\Phi_W$ is the phase space element for $W^+\to l^+\nu_l$:
\be
d\Phi_W&=&\frac{E_l^2}{2m_W^2}~
\frac{d\Omega_l}{(2\pi )^2}~,\quad
E_l=\frac{m_W^2}{2[E_t+|{\bf p}_t| c_l-E_b(1-c_{bl})]}~,
\label{PhW}
\ee
where $c_b= \cos\theta_b$, $c_l= \cos\theta_l$ and
$c_{bl}=\cos\theta_{bl}$, with $\theta_{bl}$
being the angle between ${\bf p}_b$ and ${\bf p}_l$, and
$d\Omega_b=\sin\theta_b d\theta_b d\phi_b$ etc.

\noi
\underline{\it Phase space element for $\ti\chi^-_j$ decay via $\ti\nu$ exchange (\ref{I}):}

\noi
The phase space element of the decay
$\ti\chi^-_j\to \ell^-_1 \bar{\ti\nu}$ reads
\be
d\Phi^1_{\chi_j}&=&\frac{E_{\ell_1}^2}{2(m_{\chi_j}^2-m_{\sn}^2)}~
\frac{d\Omega_{\ell_1}}{(2\pi )^2}~,
\quad  E_{\ell_1}=\frac{m^2_{\chi_j}-m^2_{\ti\nu}}
{2(E_{\chi_j}-|{\bf p}_{\chi_j}|~c_1)}~,
\label{eq:phi1}
\ee
where $c_1= \cos\theta_{\ell_1}$.

\noi
\underline{\it Phase space elements for $\ti\chi^-_j$ decay
via $\ti\ell$ exchange (\ref{II}):}

\noi
The phase space element of the decay $\ti\chi^-_j\to \ti\ell^-_n \bar{\nu}$ is given by
\be
d\Phi^{2}_{\chi_j}=
\frac{E_\nu^2}{2(m_{\chi_j}^2-m_{\sl}^2)}\,\,\frac{d\Omega_\nu}{(2\pi )^2}~,
\qquad E_\nu= \frac{m^2_{\chi_j}-m^2_{\sl}}{2(E_{\chi_j}-|{\bf p}_{\chi_j}|~c_\nu)}~,
\label{eq:phasespchiW2}
\ee
where $c_\nu=\cos\theta_\nu$.
For the subsequent decay $\ti\ell^-_n\to \ti\chi^0_1\ell^-_2$ the phase space element reads
\be
d\Phi_{\sl} &=&
\frac{E_{\ell_2}^2}{2(m_{\sl}^2-m_{\chi^0_1}^2)}~\frac{d\Omega_{\ell_2}}{(2\pi )^2}~,
\qquad E_{\ell_2}= \frac{m_{\sl}^2-m_{\chi_1^0}^2}{2(E_{\sl}-|{\bf p}_{\sl}|~c_{\sl\ell_2})}~,
\label{eq:phasespsl}
\ee
where $c_{\sl\ell_2}=\cos\theta_{\sl\ell_2}$ being
the angle between ${\bf p}_{\sl}$ and ${\bf p}_{\ell_2}$.

\noi
\underline{\it Phase space elements for $\ti\chi^-_j$ decay via $W$ boson exchange (\ref{III}):}

\noi
The phase space element of the decay $\ti\chi^-_j\to W^- \ti\chi^0_1$ is given by
\bee{eq:PchjW}
(d\Phi_{\chi_j}^3)^\pm=
\frac{|{\bf p}^{\pm}_W|^2}{4 |E_W^{\pm}~|{\bf p}_{\chi_j}|\cos\theta_W-
E_{\chi_j}~|{\bf p}^{\pm}_W||}\,\frac{d\Omega_W}{(2\pi)^2}~,
\eee
with
\be
|{\bf p}^{\pm}_W|&=&
\left[(m^2_{\chi_j}+m^2_W-m^2_{\chi^0_1})|{\bf p}_{\chi_j}| \cos\theta_W)
\right.
\nonumber \\
& \pm & {}\left.
E_{\chi_j}\sqrt{\lambda(m^2_{\chi_j},m^2_W,m^2_{\chi^0_1})-
4|{\bf p}_{\chi_j}|^2~m^2_W~(1-\cos^2\theta_W)}
\right]
\nonumber \\
& \times & {}
\left[2|{\bf p}_{\chi_j}|^2 (1-\cos^2\theta_W)+2 m^2_{\chi_j})
\right]^{-1}~.
\ee
There are two solutions $|{\bf p}^{\pm}_W|$ in the case
$|{\bf p}^0_{\chi_j}|<|{\bf p}_{\chi_j}|$,
where $|{\bf p}^0_{\chi_j}|=$
$\frac{\sqrt{\lambda(m^2_{\chi_j},m^2_W,m^2_{\chi^0_1})}}{2m_W}$
is the chargino momentum if the $W$ boson is produced at rest.
The $W$ decay angle $\theta_W$ is constrained in that case
and the maximal angle $\theta^{\rm max}_W$ is given as
\be
\sin\theta^{\rm max}_W=
\frac{|{\bf p}^0_{\chi_j}| }{|{\bf p}_{\chi_j}| }=
\frac{m_{\ti b_m}}{m_W}
\frac{\lambda^{\frac{1}{2}}(m^2_{\chi_j},m^2_W,m^2_{\chi^0_1})}
{\lambda^{\frac{1}{2}}(m^2_{\ti b},m^2_{\chi_j},m^2_t)}\leq 1~.
\ee
If $|{\bf p}^0_{\chi_j}|>|{\bf p}_{\chi_j}|$, the
decay angle $\theta_W$ is not constrained and there is
only the physical solution $|{\bf p}^+_W|$.

For the subsequent decay of the $W$ boson, $W^-\to \ell^-_3\nu$, the phase space element
is analogous to the one given in (\ref{PhW}) and reads
\be
d\Phi_W^3= \frac{E_{\ell_3}^2}{2m_W^2} \frac{d\Omega_{\ell_3}}{(2\pi )^2}~,
\qquad
E_{\ell_3}=
\frac{m_W^2}{2(E^\pm_W-|{\bf p}^\pm_W|~c_{\ell_3 W})}~,
\label{PhW3}
\ee
where $c_{\ell_3 W}=\cos\theta_{\ell_3 W}$ being the angle between
${\bf p}_{\ell_3}$ and ${\bf p}_W$.

\end{appendix}

\end{document}